\documentclass[aps,prb,reprint]{revtex4-2}
\usepackage{graphicx}   
\usepackage{amsmath,amssymb,amsfonts}
\usepackage{physics}
\begin{document}

\title{Contribution of electric self-forces to electromagnetic momentum in a moving system} 
\author{Ashok K. Singal}
\email{ashokkumar.singal@gmail.com}
\affiliation{Astronomy and Astrophysics Division, Physical Research Laboratory,
Navrangpura, Ahmedabad - 380 009, India }
\date{\today}

\begin{abstract}
In moving electromagnetic systems, electromagnetic momentum calculated from the vector potential is shown to be proportional to the field energy of the system. The momentum thus obtained is shown actually to be the same as derived from a Lorentz transformation of the rest-frame electromagnetic energy of the system, assuming electromagnetic energy-momentum to be a 4-vector. The energy-momentum densities of electromagnetic fields form, however, components of the electromagnetic stress-energy tensor, and their transformations from rest frame to another frame involve additional contributions from stress terms in the Maxwell stress tensor which do not get represented in the momentum calculated from the vector potential. The genesis of these additional contributions, arising from stress in the electromagnetic fields, can be traced, from a physical perspective, to electric self-forces contributing to the electromagnetic momentum of moving systems that might not always be very obvious. Such subtle contributions to the electromagnetic momentum from stress in electromagnetic fields that could be significant even for non-relativistic motion of the system. Such contributions from stress in electromagnetic fields also provide a natural solution to some curious riddles in electromagnetic momentum like the famous, century-old, enigmatic factor of 4/3, encountered in the electromagnetic momentum of a moving charged sphere.
\end{abstract}

%

\maketitle
\section{Introduction}
There is an apparent paradox in the field electromagnetic momentum of a moving charged parallel-plate capacitor. One expects such a system to possess, along the direction of motion, an electromagnetic momentum that can be calculated from the electromagnetic field energy within the moving capacitor. However, for the capacitor plate surfaces parallel to the direction of motion, electromagnetic momentum computed from its fields turns out to be twice that expected from the field energy of the system, while for plate surfaces perpendicular to the direction of motion, the field momentum computation yields a nil value \cite{15,31}. One could, instead, use an alternate formulation, based on vector potential, to calculate the electromagnetic momentum of the moving charged parallel-plate capacitor, which yields a result independent of the plate orientations \cite{On22}. The electromagnetic momentum computed that way, however, does not yield the {\em total electromagnetic momentum} of the system, as computed from the standard definition of the electromagnetic momentum density, there seem to some contribution to the electromagnetic momentum missing. 

Elsewhere, it has recently been shown that additional contributions to momentum  
arise from the pressure or stress within a perfect fluid, that could be comprising an ideal gas, when the system is undergoing a bulk motion and similar is the case in electromagnetic systems \cite{84,85}. From that it could accordingly be construed that it is the stress, arising due to the electromagnetic forces between various constituents of the charged system,  whose contribution in a moving electromagnetic system might apparently be the term that is missing in electromagnetic momentum computed from vector potential. 
However, doubts have been expressed \cite{On22} not only upon whether the arguments employed in the case of perfect fluids \cite{84} are really compatible with and could thus be applied in electromagnetic systems, but even on whether the standard definition of the electromagnetic momentum density itself is proper and thus might need to be modified. 

Here, we examine such doubts in detail and provide simple arguments to dispel them, clarifying the basic physics involved. For that purpose, we compute the electromagnetic momentum of a moving system from the electromagnetic stress-energy tensor, determining contributions arising from various stress terms in the Maxwell stress tensor. We show that these additional contributions could be significant even for a non-relativistic motion of the system. It should be clarified that even though we shall be dealing with non-relativistic motion of the system, all our arguments are under the aegis of the special theory of relativity, in particular the mass-energy relation, where in an electromagnetic system, the electric potential energy of the charges (or the equivalent electric field energy) contributes to the mass of the system and that there is momentum associated with the energy flux in a system. We thereby demonstrate that even without referring to the fluid description, one arrives at similar results using the stress energy tensor in electromagnetic systems. 
Although we do not explicitly invoke the fluid picture here, the underlying physics and the mathematics involved are essentially the same in the either description since pressure in a fluid system corresponds to stress in the electromagnetic fields, and this correspondence between the two does add to the physical perspective. 

The electromagnetic energy-momentum density can be computed 
from the symmetric, trace-free, electromagnetic stress-energy tensor \cite{MTW73} (in cgs units)
\begin{eqnarray}
\label{eq:84.223}
{T}^{\mu\nu} = \frac{1}{4\pi} [ F^{\mu\alpha}\:\eta^{\nu\xi}F_{\xi\alpha} - \frac{1}{4} \eta^{\mu\nu}F^{\alpha\xi}F_{\alpha\xi}] \,,
\end{eqnarray}
where $F^{\mu \nu }$ and $F_{\mu \nu }$ stand for the antisymmetric electromagnetic field tensors 
\begin{eqnarray}
\nonumber
F^{\mu \nu}
&=&{\begin{bmatrix}0&E_{\rm x}&E_{\rm y}&E_{\rm z}\\-E_{\rm x}&0&B_{\rm z}&-B_{\rm y}\\-E_{\rm y}&-B_{\rm z}&0&B_{\rm x}\\-E_{\rm z}&B_{\rm y}&-B_{\rm x}&0\end{bmatrix}},\\
\label{eq:84.22}
F_{\mu \nu }
&=&{\begin{bmatrix}0&-E_{\rm x}&-E_{\rm y}&-E_{\rm z}\\E_{\rm x}&0&B_{\rm z}&-B_{\rm y}\\E_{\rm y}&-B_{\rm z}&0&B_{\rm x}\\E_{\rm z}&B_{\rm y}&-B_{\rm x}&0\end{bmatrix}}\,,
\end{eqnarray}
and $\eta^{\mu \nu }(=\eta_{\mu \nu})$ is the metric tensor of the special relativity \cite{MTW73,SC85}
\begin{eqnarray}
\label{eq:84.2}
{\displaystyle \eta^{\mu \nu }
{\displaystyle ={\begin{bmatrix}-1&0&0&0\\0&1&0&0\\0&0&1&0\\0&0&0&1\end{bmatrix}}} 
 =\operatorname {diag} (-1,1,1,1)}\:.
\end{eqnarray} 
We follow the convention where Greek letters $\alpha, \mu, \nu, \xi$ etc. take the values $0,1,2,3$ while Latin letters $i,j$ take the values $1,2,3$. 

From the stress-energy tensor (Eq.~(\ref{eq:84.223})), using Eqs.~(\ref{eq:84.22}) and (\ref{eq:84.2}), we get the electromagnetic energy density  
\begin{eqnarray}
\label{eq:84a.24}
{T}^{00}={\frac {1}{8\pi }}(E^{2}+B^{2})\,,
\end{eqnarray}
and the energy flux ($c{T}^{0i}$) in $i$th direction, as well as $\it i$th component of the electromagnetic momentum density (${T}^{i0}/c$) as 
\begin{eqnarray}
\label{eq:84a.25}
{T}^{i0} = {T}^{0i} ={\frac {1}{4\pi}}(\mathbf {E} \times \mathbf {B})^i\,. 
\end{eqnarray}
while the symmetric Maxwell stress tensor, given by
\begin{eqnarray}
\label{eq:84a.26}
{\displaystyle {T}^{ij}=\frac {1}{8\pi }\left[\left(E^{2}+B^{2}\right)\delta^{ij}-2(E^{i}E^{j}+B^{i}B^{j})\right]}
\end{eqnarray}
denotes $\it i$th component of the momentum flux in $\it j$th direction. Kronecker delta, $\delta^{ij}=1$ for $i=j$ and zero otherwise. 
Even in the simplest of electromagnetic field configurations, for instance for a uniform electric field in a region, as in the case of a charged parallel plate capacitor, there is finite stress in the fields. From the Maxwell stress tensor (Eq.~(\ref{eq:84a.26})) we find a tension or negative pressure, $-(E^{2}+B^{2})/{8\pi }$, along the field lines, however, at right angles to the field lines there is a positive pressure \cite{MTW73,RI06}.

We want to determine the electromagnetic momentum of a system in the lab frame $\cal K$, where the rest frame $\cal K'$ of the system is moving with a uniform velocity $\mathbf v$ along the $x$-axis. With $\beta=v/c$ and the Lorentz factor $\gamma =1/\sqrt{1-\beta^{2}}$, matrix 
for Lorentz transformation,  
${\displaystyle  x^{\mu}=\Lambda^{\mu}_{\alpha'}x^{\alpha'}}$,
between coordinates $x^{\mu}\equiv [x^0,x^1,x^2,x^3]=[ct,x,y,z]$ of $\cal K$ and $x^{\alpha'}$ of $\cal K'$ is given by \cite{MTW73}  
\begin{eqnarray}
\label{eq:84.1a}
{\displaystyle \Lambda^{\mu}_{\alpha'}=
 {\begin{bmatrix}\gamma&\gamma \beta&0&0\\\gamma \beta&\gamma&0&0\\0&0&1&0\\0&0&0&1\end{bmatrix}}}\:,
\end{eqnarray}
a prime on a quantity indicating it refers to the rest frame  $\cal K'$ of the charged system.
A Lorentz transformation of the stress-energy tensor from the rest frame  $\cal K'$, where the electromagnetic momentum may be taken as zero, to the lab frame $\cal K$ is then given by 
\begin{eqnarray}
\label{eq:84.1b}
{\displaystyle {T}^{\mu\nu}={\Lambda ^{\mu}}_{\alpha'}{\Lambda ^{\nu}}_{\xi'}T'^{\alpha' \xi'}\,.} 
\end{eqnarray}
Throughout, we denote the stress-energy tensor in the rest frame with a prime, $T'$, in order to distinguish its components from those in the lab frame, for instance, $ T^{11}$ represents one of the components of the Maxwell stress tensor in the lab frame while $ T'^{11}$ stands for the corresponding component in the rest frame. It will be shown that in the energy flux ($c{T}^{0i}$) as well as the electromagnetic momentum density (${T}^{i0}/c$) in the lab frame there are terms arising from the contribution of Maxwell stress tensor ($ T'^{i'j'}$) in the rest frame. It is these additional terms  which lead to the resolutions of the above-mentioned paradox in the moving charged capacitor as well as an explanation of the mysterious factor of 4/3 in the electromagnetic momentum of a moving charge, a puzzle that apparently defied a proper resolution for almost a century \cite{1,2,29}. 
\section{Alternate definitions of electromagnetic momentum of a charged system}
\subsection{Electromagnetic momentum of a charged system from vector potential}
Electromagnetic momentum in the fields of a charged parallel-plate capacitor, computed from the volume integral of Eq.~(\ref{eq:84a.25})
\begin{eqnarray}
\label{eq:84a.4}
{\mathbf P}_{\rm f}&=&\frac {1}{4\pi c}\int(\mathbf {E} \times \mathbf {B})\,{\rm d}\tau\,,
\end{eqnarray}
gives a finite value for the capacitor moving parallel to the plate surfaces, however, it yields a {\em nil} value (with ${\mathbf B}=0$) for the same capacitor with a motion normal to the plate surfaces \cite{15}.
%

Alternatively, taking a cue from the generalized force on a point charge $q$ due to velocity-dependent  electromagnetic potentials \cite{Go50,Da20}
\begin{eqnarray}
\label{eq:84a.3}
\frac{\rm d}{{\rm d}t}\Big[m {\mathbf v}+\frac{q{\mathbf A}}{c}\Big]&=& -q\grad\Big[\phi-\frac{{\mathbf v}\cdot {\mathbf A}}{c}\Big]\,,
\end{eqnarray}
where $m{\mathbf v}$ is the mechanical momentum of the point charge moving with an instantaneous non-relativistic velocity ${\mathbf v}$, and $\phi$ and ${\mathbf A}$ are the scalar and vector potentials at the location of the charge $q$, 
the electromagnetic momentum of a point charge $q$ could be
\begin{eqnarray}
\label{eq:84a.4d1}
{\mathbf P}_{\rm A}&=&\frac {q{\mathbf A}}{c}\,.
\end{eqnarray}
This expression for electromagnetic momentum associated with a point charge $q$ was suggested by Maxwell \cite{Ma65}. This form of electromagnetic momentum has been discussed at some length in the literature \cite{Ko78,Se98,Gr12,Es18}.

The scalar and Vector potentials $\phi$ and ${\mathbf A}$, for continuous charge density $\rho$ and current density $\mathbf j=\rho {\mathbf v}$ distributions, are computed from the volume integrals
\begin{eqnarray}
	\label{eq:84a.4b2}
	\phi({\mathbf x})&=&\int\frac {\rho({\mathbf x}'))}{|{\mathbf x}-{\mathbf x}'|}\,{\rm d}\tau'\\
	\label{eq:84a.4b}
	{\mathbf A}({\mathbf x})&=&\int\frac {{\mathbf j}({\mathbf x}')}{c|{\mathbf x}-{\mathbf x}'|}\,{\rm d}\tau=\int\frac {\rho({\mathbf x}'){\mathbf v}({\mathbf x}')}{c|{\mathbf x}-{\mathbf x}'|}\,{\rm d}\tau'\,,
\end{eqnarray}
with $|{\mathbf x}-{\mathbf x}'|$ as the distance of the charge element $\rho({\mathbf x}')\,{\rm d}\tau'$, that may be moving with velocity $\mathbf v({\mathbf x}')$, from the field point ${\mathbf x}$ where $\phi$ and $\mathbf A$ are to be evaluated. The expression for $\phi$ and ${\mathbf A}$ (Eqs.~(\ref{eq:84a.4b2}) and (\ref{eq:84a.4b})) are for the Lorenz gauge, but without the retarded time condition. This is because we are dealing  with non-relativistic velocities, where even otherwise, a specification of the gauge may not be of much importance.
For a discrete charge $q$, $q \phi$ denotes electrostatic potential energy of the charge, while $q A/c$ is supposed to represent electromagnetic momentum associated with the charge. If the whole system represented by the charge density $\rho$ is moving with a common velocity vector ${\mathbf v}$, then from Eqs.~(\ref{eq:84a.4b2}) and (\ref{eq:84a.4b}) ${\mathbf A}=\phi{\mathbf v}/c$, and in that case the electromagnetic momentum of the charge $q$ is given by $q\phi{\mathbf v}/c^2$.

In the case of $N$ discrete charges $q_{\rm j}$, with velocity vectors ${\mathbf v}_{\rm j}$, the vector potential ${\mathbf A}$ at the  location of the charge $q$ is computed from the summation
\begin{eqnarray}
\label{eq:84a.4e1}
{\mathbf A}({\mathbf x}_{0})&=&\frac {1}{c}\sum_{j=1}^{N} \frac {q_{\rm j}\,{\mathbf v}_{\rm j}}{|{\mathbf x}_{\rm j}-{\mathbf x}_{0}|}\,,
\end{eqnarray}
where $|{\mathbf x}_{\rm j}-{\mathbf x}_{0}|$ is the distance of charge $q_{\rm j}$, moving with velocity ${\mathbf v}_{\rm j}$, from the location ${\mathbf x}_{0}$ of the charge $q$. 

Then we have  
\begin{eqnarray}
\label{eq:84a.4e2}
{\mathbf P}_{\rm A}=\frac {q {\mathbf A}}{c}=\frac {1}{c^2}\sum_{j=1}^{N} \frac {q\, q_{\rm j}}{|{\mathbf x}_{\rm j}-{\mathbf x}_{0}|}\,{\mathbf v}_{\rm j}\,.
\end{eqnarray}

Equation~(\ref{eq:84a.4e2}) can be written in terms of the scalar potential $\phi_{\rm j}=q/|{\mathbf x}_{\rm j}-{\mathbf x}_{0}|$, due to $q$ at the location ${\mathbf x}_{\rm j}$ of charge $q_{\rm j}$, as 
\begin{eqnarray}
\label{eq:84a.4e3}
{\mathbf P}_{\rm A}=\frac {q {\mathbf A}}{c}=\sum_{j=1}^{N} \frac {\phi_{\rm j}\:q_{\rm j}}{c^2}\,{\mathbf v}_{\rm j}\,.
\end{eqnarray}

We can use the energy-mass relation, to express the potential energy $q_{\rm j}\,\phi_{\rm j}$ of a charge $q_{\rm j}$, owing to the presence of charge $q$ at ${\mathbf x}_0$, in terms of its mass equivalent
\begin{eqnarray}
\label{eq:84a.4d4}
\Delta m_{\rm j}=\frac {q_{\rm j}\,\phi_{\rm j}}{c^2} \,,
\end{eqnarray}
to write
\begin{eqnarray}
\label{eq:84a.4e4}
{\mathbf P}_{\rm A}=\frac {q {\mathbf A}}{c}=\sum_{j=1}^{N} {\Delta m_{\rm j}\,{\mathbf v}_{\rm j}}
\,,
\end{eqnarray}
the right hand side is readily recognized as momentum. 
The term $\Delta m_{\rm j}$ here has nothing to do with inertial mass $m_{\rm j}$ of the $j$th charged particle and that the electromagnetic momentum in Eq.~(\ref{eq:84a.4e4}) is not sum of the kinetic momentum, $\Sigma m_{\rm j}\,{\mathbf v}_{\rm j}$, of moving charged particles. 

The contribution to the electromagnetic momentum of the system from an element $\rho {\rm d}\tau$ of the continuous charge distribution, instead of the discrete charge $q$, can be written as
\begin{eqnarray}
	\label{eq:84a.4a}
	{\rm d}{\mathbf P}_{\rm A}&=&\frac {1}{c}\rho({\mathbf x}){\mathbf A}({\mathbf x})\,{\rm d}\tau\,.
\end{eqnarray}
Electromagnetic momentum, ${\mathbf P}_{\rm A}$, of the total system could then be obtained by an integration over the whole charge distribution. 

Applying Eq.(\ref{eq:84a.4a}), along with Eq.(\ref{eq:84a.4b}), to a moving charged capacitor case it has been purportedly claimed \cite{On22} that the electromagnetic momentum of the system does not depend upon the plate orientations with respect to the direction of motion and a conclusion has been drawn therefrom that the fluid concept used in literature \cite{84} cannot be applied to the analysis of electrodynamical systems. 

Our contention here is that Eq.~(\ref{eq:84a.4}) actually is a much more general expression, valid even when there are no electric charges in the system, for instance it can be employed to calculate the momentum carried by an electromagnetic wave \cite{Es18}.  Further, as we shall be showing, the electromagnetic momentum computed using vector potential $\mathbf A$, is  already being taken into account and forms a part of the total electromagnetic momentum calculated using Eq.~(\ref{eq:84a.4}).
We shall further show that if in an electromagnetic system there are finite stress terms in the Maxwell stress tensor or equivalently net electromagnetic forces are present in the system (even when these might be balanced by some mechanical forces), then when the system is in uniform motion there may be electromagnetic energy exchange taking place between different sections of the system, even if there is no change in the total energy of the system. In such a case there will be additional electromagnetic momentum associated with this flow of energy within the system, and it is this additional electromagnetic momentum that represents the difference in the formulations  Eqs.(\ref{eq:84a.4}) and (\ref{eq:84a.4d1}) for the electromagnetic momentum.

The total potential energy of the system is given by 
\begin{eqnarray}
	\label{eq:84a.4c2}
{\cal E}=\frac {1}{2}\int\rho({\mathbf x})\phi({\mathbf x}){\rm d}\tau=\int\frac{E^2({\mathbf x})}{8\pi}{\rm d}\tau\, 
\end{eqnarray}
where a factor of 2 in the denominator occurs because each pair of charge element appears twice in the integration while  
${E^2}/{8\pi}$ is the electrostatic field energy density of the system \cite{1}. 

The case we are particularly interested in here is where all charges in the  system are moving with the same velocity vector, ${\mathbf v}$. In that case, the total electromagnetic momentum of the system can be written as
\begin{eqnarray}
	\label{eq:84a.4c}
	{\mathbf P}_{\rm A}&=&\frac {\mathbf v}{2c^2}\int\int\frac{\rho({\mathbf x})\rho({\mathbf x}')}{|{\mathbf x}-{\mathbf x}'|}{\rm d}\tau'{\rm d}\tau\,,
\end{eqnarray}
Using Eq.~(\ref{eq:84a.4b2}) for the scalar potential in  Eq.~(\ref{eq:84a.4c}), 
we can write 
\begin{eqnarray}
	\label{eq:84a.4c3}
		{\mathbf P}_{\rm A}&=&\frac {\mathbf v}{2c^2}\int\rho({\mathbf x})\phi({\mathbf x}){\rm d}\tau=\frac {{\cal E}{\mathbf v}}{c^2}	\,,
\end{eqnarray}
which can be written as 
\begin{eqnarray}
	\label{eq:84a.4c4}
	{\mathbf P}_{\rm A}&=&M_{\rm em}{\mathbf v}\,,
\end{eqnarray}
where $M_{\rm em}={{\cal E}}/{c^2}$ is the electromagnetic mass equivalent of the field energy of the system. 
\subsection{Electromagnetic momentum of a moving charged system computed from electromagnetic stress-energy tensor}
The electromagnetic momentum of a charged system can be determined in the lab frame from electromagnetic stress-energy tensor. For this  we do a Lorentz transformation of the stress-energy tensor from the rest frame $\cal K'$ to the lab frame $\cal K$.
Taking the energy flux and the electromagnetic momentum to be zero in rest frame $\cal K'$, the stress-energy tensor can be written as  
\begin{eqnarray}
\label{eq:84.3a0}
{\displaystyle T'^{\mu' \nu' }
{\displaystyle ={\begin{bmatrix}T'^{00}&0&0&0\\0&T'^{11}&T'^{12}&T'^{13}\\0&T'^{21}&T'^{22}&T'^{23}\\0&T'^{31}&T'^{32}&T'^{33}\end{bmatrix}}}}\:.
\end{eqnarray} 
Assuming $\cal K'$ to be moving with respect to $\cal K$ along the $x$-axis with a uniform velocity $v=\beta c$, applying Eqs.~(\ref{eq:84.1a}) and (\ref{eq:84.1b}) for a Lorentz transformation, after somewhat tedious, albeit straightforward computations, we can get components of the stress-energy tensor in frame $\cal K$, for instance, 
\begin{eqnarray}
\label{eq:84.3a111}
T^{00}&=&{\Lambda ^{0}}_{\alpha'}{\Lambda ^{0}}_{\xi'}T'^{\alpha' \xi'}={\Lambda ^{0}}_{0}{\Lambda ^{0}}_{0}T'^{00}+{\Lambda ^{0}}_{1}{\Lambda ^{0}}_{0}T'^{10}\nonumber\\
&&+{\Lambda ^{0}}_{0}{\Lambda ^{0}}_{1}T'^{01}
+{\Lambda ^{0}}_{1}{\Lambda ^{0}}_{1}T'^{11}\nonumber\\
&=&\gamma^2T'^{00}+\gamma^2\beta^2T'^{11}\,, 
\end{eqnarray} 
or
\begin{eqnarray}
\label{eq:84.3a112}
T^{01}&=&{\Lambda ^{0}}_{\alpha'}{\Lambda ^{1}}_{\xi'}T'^{\alpha' \xi'}={\Lambda ^{0}}_{0}{\Lambda ^{1}}_{0}T'^{00}+{\Lambda ^{0}}_{1}{\Lambda ^{1}}_{0}T'^{10}\nonumber\\
&&+{\Lambda ^{0}}_{0}{\Lambda ^{1}}_{1}T'^{01}
+{\Lambda ^{0}}_{1}{\Lambda ^{1}}_{1}T'^{11}\nonumber\\
&=&\gamma^2\beta T'^{00}+\gamma^2\beta T'^{11}\,. 
\end{eqnarray} 
Proceeding in this manner, we get
\footnotesize
\begin{eqnarray}
\label{eq:84.3a11}
T^{\mu \nu}&=&
{\displaystyle {\begin{bmatrix}\gamma^2(T'^{00}+T'^{11}\beta^2) &\!\!\!\!\gamma^2(T'^{00}+T'^{11})\beta &\!\! \gamma T'^{12}\beta &\!\! \gamma T'^{13}\beta\\\gamma^2(T'^{00}+T'^{11})\beta &\!\!\!\!\gamma^2(T'^{00}\beta^2+T'^{11}) &\!\! \gamma T'^{12} &\!\! \gamma T'^{13}\\
\gamma T'^{12}\beta&\gamma T'^{21}&\!\!T'^{22}&T'^{23}\\\gamma T'^{13}\beta&\!\!\gamma T'^{31}&\!\!T'^{32}&\!\!T'^{33}\end{bmatrix}}}\,.
\nonumber\\
&&
\end{eqnarray} 
\normalsize
On the right hand side in Eq.~(\ref{eq:84.3a11}), the values for the stress-energy components in the matrix are for the rest frame $\cal K'$. 

We want to evaluate $T^{\mu \nu}$ in a frame moving with a non-relativistic velocity. Then, for low velocities ($\beta=v/c\ll 1$), keeping only the lowest order terms in $\beta$ and thereby  putting $\gamma =1/\sqrt{1-\beta^{2}}\approx 1$ and dropping all terms $\propto \beta^2$ in Eq.~(\ref{eq:84.3a11}), the stress-energy tensor in lab frame $\cal K$, takes a much simpler form
\footnotesize
\begin{equation}
\label{eq:84.3a1}
T^{\mu \nu}= 
{\displaystyle {\begin{bmatrix}T'^{00}&(T'^{00}+T'^{11})\beta&T'^{12}\beta&T'^{13}\beta\\(T'^{00}+T'^{11})\beta&T'^{11}&T'^{12}&T'^{13}\\T'^{12}\beta&T'^{21}&T'^{22}&T'^{23}\\T'^{13}\beta&T'^{31}&T'^{32}&T'^{33}\end{bmatrix}}}\,.
\end{equation} 
\normalsize

Incidentally, for a transformation from the rest frame $\cal K'$ to another frame $\cal K$, moving with a non-relativistic velocity, one generally performs a Galilean transformation for time and space coordinates (in appropriate units), employing the matrix
\begin{eqnarray}
\label{eq:84.1c}
{\displaystyle \Lambda^{\mu}_{\alpha'}
  = {\begin{bmatrix}1&0&0&0\\\beta&1&0&0\\0&0&1&0\\0&0&0&1\end{bmatrix}}}\:.
\end{eqnarray}

The stress-energy tensor in the lab frame $\cal K$ to a first order in $\beta$, as expected from a Galilean transformation matrix (Eq.~(\ref{eq:84.1c})), is
\begin{eqnarray}
\label{eq:84.1c1}
T^{\mu \nu}&=& 
{\displaystyle {\begin{bmatrix}T'^{00}&T'^{00}\beta&0&0\\T'^{00}\beta&T'^{00}\beta^2+T'^{11}&T'^{12}&T'^{13}\\0&T'^{21}&T'^{22}&T'^{23}\\0&T'^{31}&T'^{32}&T'^{33}\end{bmatrix}}}\nonumber\\
&\approx& 
{\displaystyle {\begin{bmatrix}T'^{00}&T'^{00}\beta&0&0\\T'^{00}\beta&T'^{11}&T'^{12}&T'^{13}\\9&T'^{21}&T'^{22}&T'^{23}\\0&T'^{31}&T'^{32}&T'^{33}\end{bmatrix}}}\,.
\end{eqnarray} 
which is not the same as Eq.~(\ref{eq:84.3a1}).

Actually, in the present case we are using relativistic formulations like $M={\cal E}/c^2$ and $P={\cal E}v/c^2$, where ${\cal E}$ is the electromagnetic energy in fields. Moreover, the classical electromagnetic theory is known to be in conformity with the special theory of relativity. In fact, the Lorentz transformations for the 
electromagnetic fields were derived \cite{Lo04} even before  Einstein put forward the special theory of relativity \cite{Ei05}. Therefore, while dealing with the electromagnetic stress-energy tensor, to be consistent, we should instead employ Lorentz transformation (Eq. (7)), which for low velocities, keeping only the lowest order terms in $\beta$ with $\gamma =1/\sqrt{1-\beta^{2}}\approx 1$, 
becomes    
\begin{eqnarray}
\label{eq:84.1a1}
{\displaystyle \Lambda^{\mu}_{\alpha'}
  \approx {\begin{bmatrix}1& \beta&0&0\\\beta&1&0&0\\0&0&1&0\\0&0&0&1\end{bmatrix}}}\:.
\end{eqnarray}
It should be noted that by keeping only the lowest order terms in $\beta$ we are considering a non-relativistic case and not a {\em semi-relativistic} case, where usually terms up to $\beta^2$, with $\gamma \approx 1+\beta^2/2$, are retained. Using the transformation matrix (Eq.~(\ref{eq:84.1a1})), we get back Eq.~\ref{eq:84.3a1} as the expression for the stress-energy tensor in the lab frame $\cal K$.


A comparison of Eqs.~(\ref{eq:84.3a0}) and (\ref{eq:84.3a1}) shows that in the stress-energy tensor, when transformed to the lab frame for non-relativistic velocities, there is no change in the energy density $T^{00}$ as well as in Maxwell stress tensor  $T^{ij}$. However, $T'^{00}$ and $T'^{1i'}$ in the rest frame make contributions to the energy flux $cT^{0i}$ and momentum density $T^{i0}/c$ in the lab frame, even for a non-relativistic motion of the system along $x$-axis.
There appears thus in energy flux and momentum density each, a term $\propto T'^{00}v$, due to the system carrying stored energy density $T'^{00}$ with a uniform velocity $v$, while additional terms $\propto T'^{1i'}v$ in the energy flux and momentum density in the lab frame arise owing to stress $T'^{1i'}$ present in the fields in the rest frame. As will be shown further it is the difference between the expressions for the energy-momentum tensor in the lab frame $\cal K$ expected from a Galilean transformation (\ref{eq:84.1c1}) versus the one obtained from a Lorentz transformation for non-relativistic velocities (\ref{eq:84.3a1}), that is responsible for the electromagnetic momentum of the system in $\cal K$ having some additional terms that apparently are quite intriguing, in particular the more than a century long, enigmatic factor of 4/3 in the electric momentum of a classical, charged particle \cite{29}.

A volume integral of the term $T'^{00}{v}/c^2$ in the momentum density $T^{i0}/c$ in Eq.~(\ref{eq:84.3a1}) yields the same value as $P_{\rm A}= {{\cal E}{v}}/{c^2}$ in Eq.~(\ref{eq:84a.4c3}) computed from the vector potential. 
In fact, that would be the correct result for the electromagnetic momentum of the system in the lab frame $\cal K$ if energy-momentum of electromagnetic fields were to transform as components of a 4-vector. However, energy-momentum densities of electromagnetic fields being components of the stress-energy tensor, their transformation from rest frame $\cal K'$ to lab frame  $\cal K$ (Eq.~(\ref{eq:84.1b})), depending upon the direction of velocity $v$, gets contributions from some terms in Maxwell stress tensor as well. Thus $P_{\rm A}$ does not, in general, give the full value of the electromagnetic momentum as it does not incorporate the additional terms ($T'^{1i'}v/c^2$ here) from stress in the fields.
Equation~(\ref{eq:84a.4d1}) might give electromagnetic momentum associated with an individual discrete charge in the presence of a vector potential, however, the derived formulation from that for the momentum of the whole consolidated system, viz. Eq.~(\ref{eq:84a.4c3}), incorporates only the term $T'^{00}v/c^2$ which is not the total electromagnetic momentum of a system, as stress-dependent terms do not get included in it.

As we shall show further, these additional contributions to the energy flux as well as to the momentum density arise from the electromagnetic forces of interaction between the constituents of the system, represented by stress in the fields. 
The same could be determined also by recognizing the contribution to the momentum of the system from the electromagnetic self-forces, which may not always be very apparent. 

Electromagnetic momentum computed from fields (Eq.~(\ref{eq:84a.4})), which for instance yields in the electromagnetic momentum of a moving charged sphere an extra factor of $4/3$  \cite{29}, may seemingly appear anomalous only when compared with the expectation from a Lorentz transformation of the rest frame field energy, treating it as one of the component of a 4-vector energy-momentum of electromagnetic fields. However, if a Lorentz transformation is carried out, instead, for the stress-energy tensor, where contributions of the stress in fields also get included, then one sees no apparent anomaly in the derived results. 

In subsequent sections, we shall apply the formulations developed here, to specific cases of moving charged systems to investigate the genesis of various terms in the electromagnetic momentum and thereby explore the relation between them in alternate approaches.
\section{A moving charged parallel-plate capacitor with plate surfaces perpendicular to the direction of motion}
We consider a charged parallel-plate capacitor at rest in reference frame $\cal K'$, with surface charge densities $+\sigma$ and $-\sigma$ on plates 1 and 2 respectively, to be moving in the lab frame $\cal K$ with a uniform velocity $\mathbf v$, which is non-relativistic ($\beta=v/c\ll 1$) so that we keep in our formulations only the lowest order terms in $\beta$. We assume the capacitor plate surfaces are perpendicular to the direction of motion. For definitiveness we assume $\mathbf v$ to be along the $x$-axis, with the capacitor plates in shape of circular discs, each of radius $L$, lying in the y--z plane. Then for $t=0$, $x'=x$. 

In order to calculate the electromagnetic momentum of a moving system from the vector potential, we begin by first computing the scalar potential $\phi'$, in the rest frame $\cal K'$, at a point between the capacitor plates that lies at a distance, say, $x'=x$ from plate 1. For this we include contributions to $\phi'$ from both the capacitor plates (Fig.~{\ref{F1}), as 
\begin{eqnarray}
\label{eq:84a.5}
 {\phi'} &=& \int_{0}^{L}\frac{\sigma 2\pi r \,{\rm d}r}{[x'^{2}+r^{2}]^{1/2}}+\int_{0}^{L} \frac{-\sigma 2\pi r\,{\rm d}r}{[(d-x')^{2}+r^{2}]^{1/2}}\nonumber\\
   &=& {\sigma 2\pi }\Big[[x'^{2}+L^{2}]^{1/2}-x'-[(d-x')^{2}+L^{2}]^{1/2}\nonumber\\
   &&+(d-x')\Big] \,,
\end{eqnarray}
where $L$ is the individual plate dimension. 

For $x'\le d\ll L$, we can write it as
\begin{eqnarray}
\label{eq:84a.6}
 {\phi'}  &=& 2\pi\sigma (d-2x') \,.
\end{eqnarray}
Thus taking $\phi'=0$ at $x'=d/2$, midway between the plates, and with the restriction $|x'|\ll L$, we get $\phi'=2\pi\sigma d$ at plate 1 ($x'= 0)$, and $\phi'=-2\pi\sigma d$ at plate 2 ($x'= d$). The vector potential ${\mathbf A}'=0$ everywhere, in the rest frame $\cal K'$.

\begin{figure}[t!](topskip=0pt, botskip=0pt, midskip=0pt)
\begin{center}
\includegraphics[width=\columnwidth]{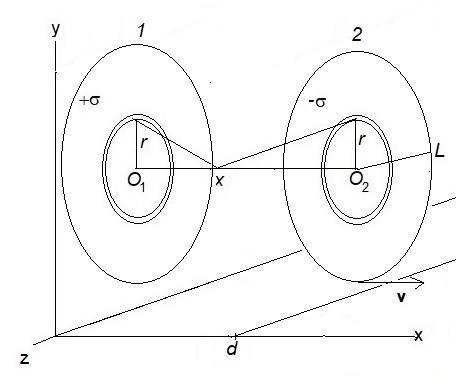}
\caption{Calculation of the scalar potential, ${\phi}$ at a point $x$ between the charged capacitor plates, due to circular rings of radii $r$, centered on $O_1$ and $O_2$. The capacitor plates, circular discs of radii $L$ are assumed to be lying in the y--z planes,  with a plate separation $d$ along the $x$ direction. The surface charge densities are $+\sigma$ and $-\sigma$, on plates 1 and 2 respectively. The capacitor is moving along the $x$-axis with a velocity $\mathbf v$, assumed to be non-relativistic.\label{F1}
}
\end{center}
\end{figure}

The electric field $\mathbf E'$ between the capacitor plates is then obtained, in frame $\cal K'$, as 
\begin{eqnarray}
\label{eq:84a.6a}
{\mathbf E'}&=& -\grad'{\phi'}  = 4\pi \sigma \hat{\mathbf x}'\,,
\end{eqnarray}
which is constant in regions between the capacitor plates. The magnetic field is zero everywhere in frame $\cal K'$.

Now we can calculate the potentials in frame $\cal K$ from a Lorentz transformation (Eq.~(\ref{eq:84.1a1})) of the 4-vector ($\phi,{\mathbf A}$) from frame $\cal K'$ (Eq.~(\ref{eq:84a.6})), at $t=0$ with $x'=x$. as
\begin{eqnarray}
\label{eq:84a.6b}
{\phi}  &=& {\phi'}= 2\pi\sigma (d-2x) \nonumber\\
A_{\rm x}&=&  {\phi'}\beta=2\pi\sigma (d-2x)\beta\,.
\end{eqnarray}
The only finite component of the vector potential ${\mathbf A}$ is along $x$-direction, with values $A_{\rm x}=2\pi\sigma v d$ at plate 1 ($x= 0)$, and $A_{\rm x}=-2\pi\sigma v d$ at plate 2 ($x= d$).

The electric field  in frame $\cal K$ is the same as ${\mathbf E'}$ in frame $\cal K'$.  
\begin{eqnarray}
\label{eq:84a.6a1}
{\mathbf E}&=& {\mathbf E'} = 4\pi \sigma \hat{\mathbf x}\,,
\end{eqnarray}
which is constant in the region between the capacitor plates. 
The magnetic field, ${\mathbf B}=\grad \times {\mathbf A}=0$ in $\cal K$ too. 

Then integrating over the charges $\sigma S$ on plate 1 and 2, where ${S}$ is the cross-section area 
on each plate, we get for the electromagnetic momentum contribution from Eq.~(\ref{eq:84a.4c3})
\begin{eqnarray}
\label{eq:84a.7}
{\mathbf P}_{\rm A}  &=& \frac{\sigma S  2\pi\sigma d-\sigma S \times(-2\pi\sigma d)}{2c^2}{\mathbf v}=\frac{\sigma S 2\pi\sigma d}{c^2}{\mathbf v} \nonumber\\
&=&\frac{2\pi \sigma^2\tau {\mathbf v}}{c^2}\,.
\end{eqnarray}
Then ${\mathbf P}_{\rm A}={{\cal E} {\mathbf v}}/{c^2}$, where ${\cal E}$ is the system potential energy equal to the electrostatic field energy in the system, given by \cite{PU85}
\begin{eqnarray}
	\label{eq:84a.7a}
	{\cal E}={2\pi\sigma^{2}\tau}=(E^2/8\pi)\tau\,.
\end{eqnarray}
Thus the electromagnetic momentum (Eq.~(\ref{eq:84a.7})), as calculated from the vector potential, depends upon the electromagnetic energy ${\cal E}$ of the system. 
This momentum is due to an overall movement of the system carrying an energy ${\cal E}$ and can be understood from the Lorentz transformation where energy ${\cal E}$ in the rest frame $\cal K'$ gives a momentum ${\cal E} {\mathbf v}/{c^2}$ in lab frame $\cal K$ for a non-relativistic motion ($\gamma \approx 1$).

However, from Eq.~(\ref{eq:84a.4}) we get in this case a conflicting value for the electromagnetic field momentum
\begin{eqnarray}
	\label{eq:84a.7c}
	{\mathbf P}_{\rm f}&=&\frac {1}{4\pi c}\int(\mathbf {E} \times \mathbf {B})\,{\rm d}\tau=0\,,
\end{eqnarray}
since in this particular case the magnetic field is zero (${\mathbf B}=0$) throughout. Thus there seems to be a mismatch in the two approaches to calculate the electromagnetic momentum of the system.

This, of course, is quite puzzling as one would have expected the two formulations to yield consistent results for the electromagnetic field momentum. Could the discrepancy be due to a choice of gauge as  Eq.~(\ref{eq:84a.4d1}) appears to be gauge dependent? Total electromagnetic momentum of the system, nevertheless, should not depend upon the chosen gauge. In any case, ${\mathbf P}_{\rm A}={\cal E} {\mathbf v}/{c^2}$, calculated from the vector potential. is giving a value, expected even on physical grounds for a system carrying an energy ${\cal E}$ and moving with velocity ${\mathbf v}$. Now as for Eq.~(\ref{eq:84a.7c}), any gauge choice cannot turn the electromagnetic field momentum zero from an otherwise finite value, as in Eq.~(\ref{eq:84a.7}). After all a gauge for the vector potential always preserves the electric and magnetic field values. So no particular gauge choice can make the magnetic field zero so as to make the electromagnetic field momentum also zero as in Eq.~(\ref{eq:84a.7c}). Therefore the discrepancy in the electromagnetic momentum values, between Eqs.~(\ref{eq:84a.7}) and (\ref{eq:84a.7c}), is due to something else that might have been overlooked and the contribution therefrom not taken into consideration.

That something is the momentum contribution, amiss from Eq.~(\ref{eq:84a.7}), that arises from the stress in the electromagnetic fields in system, given by the Maxwell stress tensor. To ascertain that missing contribution,  we first calculate  the stress-energy tensor in the rest frame $\cal K'$
\begin{eqnarray}
\label{eq:84.3a}
{\displaystyle T'^{\mu' \nu' }
 =\frac{E'^2}{8\pi}\operatorname {diag} (1,-1,1,1)}\:.
\end{eqnarray} 
The negative sign in $T'^{11}\;(=-E'^{2}/8\pi)$ here indicates tension (negative pressure) along the field lines, which are along $x$ direction  (${\mathbf E'}= E'_{\rm x} \hat{\mathbf x}$). 

It should be noted that a Galilean transformation of the the energy-momentum tensor (\ref{eq:84.1c1}) leads to Eq.~(\ref{eq:84a.7}). On the other hand, a more appropriate Lorentz transformation to the lab frame $\cal K$ for non-relativistic velocities (Eq.~(\ref{eq:84.3a1})) yields additional, off-diagonal components for the stress-energy tensor 
\begin{eqnarray}
\label{eq:84.3b}
T^{10} = T^{01} =(T'^{00} + T'^{11})\beta=0\,,
\end{eqnarray} 
where $cT^{01}$ is the energy flux in $x$ direction, and 
$T^{10}/c$ is the $x$ component of the electromagnetic momentum density. 

In Eq.~(\ref{eq:84.3b}), the term $T'^{00}\beta$ is due to the movement of the capacitor having stored energy density $T'^{00}=E^2/8\pi$, while the other term, $T'^{11}\beta$, represents the contribution of stress to the energy flux and the momentum density. The volume integral of $T'^{00}{\mathbf v}/c^2$, is the same as the momentum computed from the vector potential. The other term gives an additional contribution to momentum from stress $T'^{11}$ in fields, which as we show below, is due to the forces of attraction between the capacitor plates. Since $T'^{11} = - T'^{00}$ (Eq.~(\ref{eq:84.3a})), the net electromagnetic momentum, obtained from $T^{10}/c$ in Eq.~(\ref{eq:84.3b}) is zero, the same as in Eq.~(\ref{eq:84a.7c}).

The appearance of the stress-dependent terms in the electromagnetic momentum calculations can be seen also in the electromagnetic momentum density expression ${\mathbf E} \times {\mathbf B}/4\pi c$ in frame ${\cal K}$, where by using the relation between electric and magnetic fields, ${\mathbf B}=\mbox{\boldmath $\mathbf v$} \times {\mathbf E}/c$ \cite{PU85}, we have   
\begin{eqnarray}
\label{eq:84a.32}
\frac{1}{4\pi c}{\mathbf E} \times {\mathbf B}&=&\frac{1}{4\pi c}{\mathbf E} \times (\mbox{\boldmath $\beta$}\times {\mathbf E}) = \frac{1}{4\pi c}[\mbox{\boldmath $\beta$}E^2-{\mathbf E}(\mbox{\boldmath $\beta$} \cdot {\mathbf E})]\nonumber\\
&=&0\;,
\end{eqnarray}
where we have used $E\parallel\mbox{\boldmath $\beta$}$. Equation~(\ref{eq:84a.32}) is equivalent to  Eq.~(\ref{eq:84.3b}). 

From a physical perspective, the genesis of the additional terms in the energy flux and  momentum density, can be traced to the work done on each moving plate by the electric field caused by the opposite plate. Owing to the motion of the system, there is thus a subtle additional contribution ${\mathbf P}_{\rm add}$ to the electromagnetic momentum, arising from the mutual electric forces of attraction on the moving charged plates, a contribution that might not be so obvious. 


The capacitor plates moving with velocity ${\mathbf v}$ form two current sheets, each of thickness, say $\delta$, carrying current densities, ${\mathbf j}_1=\sigma {\mathbf v}/\delta$ on plate 1, and ${\mathbf j}_2=-\sigma {\mathbf v}/\delta$ on plate 2. The  electric field at the location of each plate, due to the other plate, is ${\mathbf E} = 2\pi \sigma \hat{\mathbf x}$ \cite{PU85}.
As a result, on plate {1} the electric field does work at a rate ${\mathbf j}_{1}\cdot {\mathbf E}$ per unit volume, while on plate {2} it is  ${\mathbf j}_{2}\cdot {\mathbf E}$. Then total work being done per unit time on plate {1} is ${\mathbf j}_{1}\cdot {\mathbf E} S\delta=2\pi\sigma^{2}S v$, while it is ${\mathbf j}_{2}\cdot {\mathbf E}S\delta=-2\pi\sigma^{2}S v$ on plate 2.
Equivalently, due to mutual force of attraction, ${\mathbf F}_1=2\pi\sigma^{2}S\,\hat{\mathbf x}$ on plate 1  and  ${\mathbf F}_2=-2\pi\sigma^{2}S\, \hat{\mathbf x}$ on plate 2 (Fig.~{\ref{F1}) \cite{PU85}, work is being done at a rate ${\mathbf F}_{1}\cdot \mathbf{v}$ on plate {1} and ${\mathbf F}_{2}\cdot \mathbf{v}$ on plate 2. 
Even though there is no net change in the system energy, plate 1 gains energy at the cost of plate 2, at a temporal rate $2\pi\sigma^{2}S v$. This implies a continuous flow of energy $2\pi\sigma^{2}S v$ per unit time, taking place across the distance of plate {1} from plate {2}, due to the electric self-fields of the moving capacitor. This transport of energy across a distance $d$ per unit time forms an additional electromagnetic momentum contribution 
\begin{equation}
\label{eq:p31.2}
{\mathbf P}_{\rm add}= -2\pi\sigma^{2}S d \frac {\mathbf v}{c^{2}} = -\frac{E^2\tau}{8\pi c^{2}} {\mathbf v}\,.
\end{equation}
to the system. 

Such a transport of energy taking place between the two capacitor plates would be quite obvious if the two plates were not constrained to stay together (by some non-electromagnetic forces!) and were thus free to move under the influence of the electric field. In that case, plate {1} would have gained a velocity increment above ${\mathbf v}$ while the velocity of plate {2} will have an equal reduction below ${\mathbf v}$. 


The momentum term ${\mathbf P}_{\rm add}$ is in addition to ${\mathbf P}_{\rm A}$, calculated in Eq.~(\ref{eq:84a.7}) using the vector potential. The total electromagnetic momentum of the system, therefore, is
\begin{equation}
\label{eq:p31.2a}
{\mathbf P}_{\rm em}={\mathbf P}_{\rm A}+ {\mathbf P}_{\rm add}=\frac{2\pi\sigma^{2}\tau}{c^{2}}{\mathbf v} - \frac{E^2\tau}{8\pi c^{2}} {\mathbf v}\,=\,0,
\end{equation}
which is consistent with the conclusion reached from Eq.~(\ref{eq:84.3b}), and is in agreement with electromagnetic momentum value being nil, as arrived at in Eq.~(\ref{eq:84a.7c}). Thus the momentum in fields actually represents the total electromagnetic momentum (${\mathbf P}_{\rm em}\equiv{\mathbf P}_{\rm f}$) of the system.

The electromagnetic momentum calculated in Eq.~(\ref{eq:84a.7}) accounts for only the first term, viz. $2\pi\sigma^{2}\tau {\mathbf v}/{c^2}$, in Eq.~(\ref{eq:p31.2a}).
The electromagnetic momentum computed from the vector potential in Eq.~(\ref{eq:84a.7}) does not seem to account for momentum in the system due to the energy transport in the moving system owing to the mutual electric force between the system constituents.
Thus we also see that the value of momentum in the electromagnetic fields of the capacitor, as calculated using Eq.~(\ref{eq:p31.2a}), which is the same as in  Eq.~(\ref{eq:84a.7c}), though derived here without any reference to the fluid picture, yet turns out to be identical to the value derived from the fluid description \cite{84}. 

\subsection{Momentum associated with energy flow due to work done by non-electromagnetic stabilizing forces}
\begin{figure}[t]
	\includegraphics[width=6.5cm]{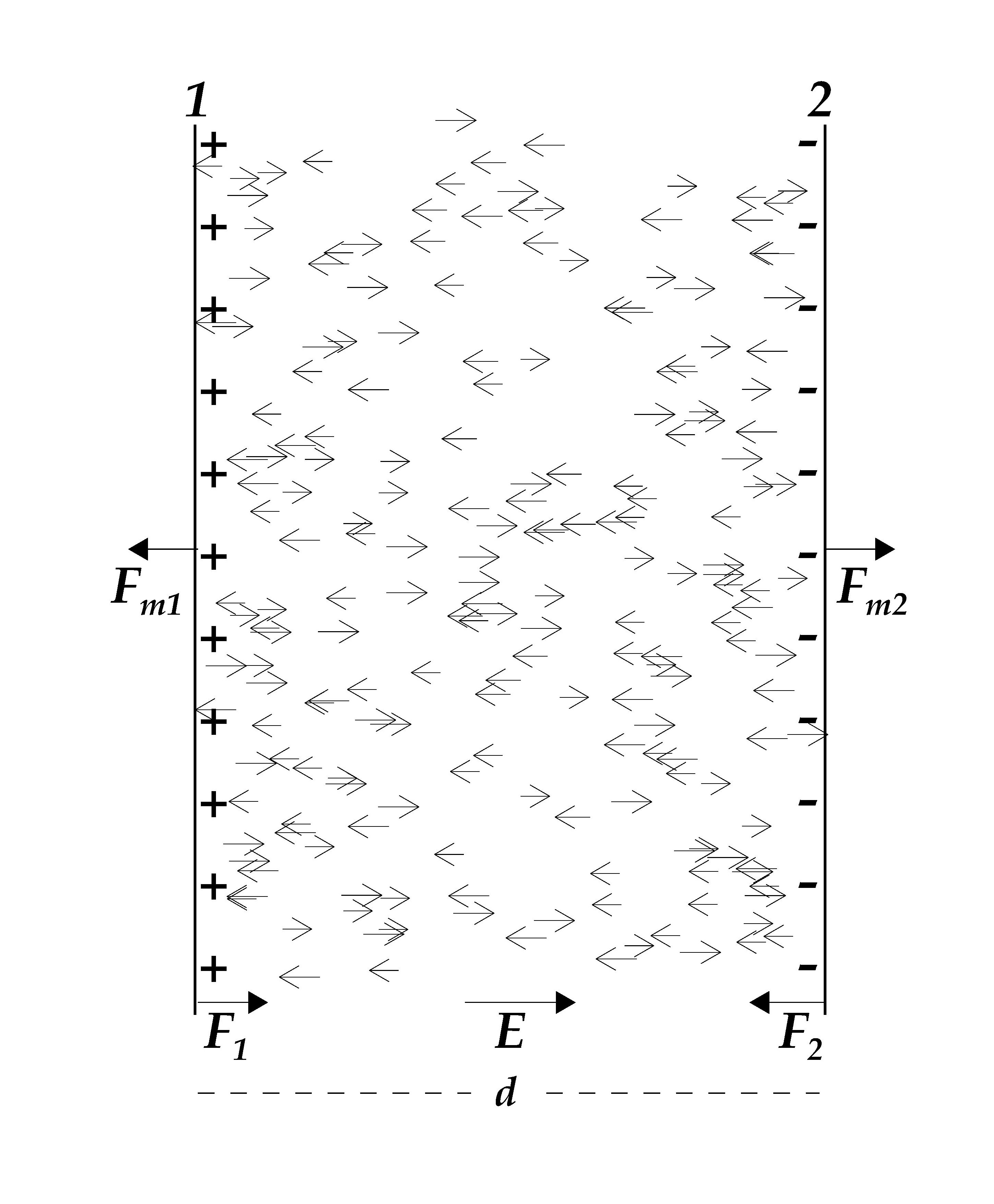}
	\caption{In the charged parallel-plate capacitor, plate {1} is experiencing an electric attractive force ${\bf F}_{1}$ along $x$-axis, while plate {2} is attracted with electric force ${\bf F}_{2}$ in the opposite direction. We assume the capacitor to contain an ideal gas comprising molecules having 1-dimensional motion with non-relativistic speeds $\pm V$ along the x-axis, the direction of plate separation, and that the resulting pressure keeps the capacitor plates separated at a   fixed distance $d$, by cancelling the electric attractive forces, through mechanical forces, ${\bf F}_{\rm m1}=-{\bf F}_{1}$ on plate {1} and ${\bf F}_{\rm m2}=-{\bf F}_{2}$ on plate {2}.\label{F2}}
\end{figure}

Since the capacitor plates continue to move with a constant velocity and their separation remains unchanged, one may ask the question where is the change in the energy of individual plates if work is being done on them by the mutual force of electric attraction. 
Actually, in such a system in equilibrium, in spite of the force of attraction between the capacitor plates, there must exist some stabilizing non-electromagnetic (!) forces (Poincar\'{e} stresses \cite{34}), e.g. the `mechanical' forces keeping the two capacitor plates apart at a fixed distance, like a rod between the plates keeping them apart or some clamps holding the plates in place. These stabilizing forces will necessarily be equal and opposite to the electric forces to keep the system in equilibrium and would thus give rise to an energy flow making equal but opposite contributions to the momentum of the moving system.  We shall demonstrate this using a simple model where the capacitor plates are
kept apart by the pressure exerted by an ideal gas whose molecules have one-dimensional motion, and the resulting pressure balances exactly the mutual force of attraction between the capacitor plates. We shall compute the work done by the gas molecules and as we will show, not only is there a transport of mechanical energy, there is also a consequential mechanical momentum due to that in the system,

We assume the capacitor to be comprising an ideal gas of a number density $n$ of molecules of mass $m$, and having 1-dimensional motion along x-axis with speeds $\pm V$ in the rest frame $\cal K'$. We assume $V$ to be non-relativistic ($V\ll c$), so we keep terms in our formulations to the lowest order in $V/c$.
We further assume the molecules of the gas to be undergoing elastic collisions with the capacitor plates, thus giving rise to a finite pressure $p$, which in turn balances the mutual  attractive force per unit area between the capacitor plates, implying $p=2\pi\sigma^{2}$.

In the rest frame ${\cal K}'$, a molecule, on encountering plate 2 of the capacitor, impinges upon it with a normal velocity $V$, rebounds with a velocity $-V$, and thereby imparts a momentum $2mV$ along $x$ direction, to the plate.
As only half of the molecules move along the $x$ axis (the other half are moving along the $-x$ axis), the number of molecules that hit a unit area of the capacitor plate 2 per unit time is $n V/2 $, giving rise to a pressure $p$, which is force per unit area, as $p=nmV^2$. Similar is the pressure on plate 1. With $S$ as the cross-section area of each plate, the molecules will be exerting a mechanical force $pS$ on each capacitor plate. These mechanical force, ${\mathbf F}_{\rm m1}$ on plate {1} is equal and opposite to ${\mathbf F}_{1}$, the force of electric attraction due to plate 2, while ${\mathbf F}_{\rm m2}$ on plate {2} is equal and opposite to ${\mathbf F}_{2}$, thereby cancelling the electric attractive force on each plate, to keep the capacitor system stable by maintaining the plate separation $d$ (Fig.~{\ref{F2}).

We assume that in the lab frame $\cal K$, the capacitor moves with a velocity $v$ ($v<V$) along  the x-axis. Then one half of the molecules will be moving in the lab frame with velocity $V+v$, while the other half of molecules will be moving in the lab frame with velocity, $-V+v$. As the capacitor plates in the lab frame, are moving towards right with a speed $v$, the molecules (moving towards right) with a velocity $V+v$ will have to catch up with the {\em receding away} plate 2 with a relative speed $|V|$, while the molecules (moving towards left) with a velocity $-V+v$, will meet plate 1 head on, again with a relative speed $|V|$. 

Now, a molecule upon reflection from plate 1, will have its energy increased from $m (V-v)^2/2$ to  $m (V+v)^2/2$, a change of $2 m V v$. With $nV/2$ as the number of molecules hitting plate 1 per unit time, a unit area of the plate will be imparting energy to molecules at a temporal rate of $mn V^2 v=pv$, and at the other end of the capacitor, the molecules bouncing from plate 2, with their energy changing from $m (V+v)^2/2$ to $m (V-v)^2/2$, will be delivering energy to plate 2 at the same rate, $pv$, per unit time. Thus while molecules will be depleting energy $pS v$ per unit time from plate 1, simultaneously at a distance $d$ away the molecules will be delivering energy  $pS v$ per unit time to plate 2. There will thus be through the system a continuous transport of energy, brought about by molecules at a rate $pSv$ per unit time, across the distance $d$ from plate {1} to plate {2}. This is equivalent to a mass transport $(p S v/c^2)$  across a distance $d$ per unit time implying for the molecular gas, a mechanical momentum 

\begin{eqnarray}
	\label{eq:p31.2b1}
	{\mathbf P}_{\rm m}&=&\frac{p S d}{c^2}{\mathbf v} = \frac{p\tau}{c^{2}}{\mathbf v}\,.
\end{eqnarray}
 which is consistent with there being a momentum  $(p \tau {\mathbf v}/ c^2)$ owing to the pressure $p$ in an ideal fluid with a non-relativistic bulk flow speed ${\mathbf v}$ \cite{85,84,MTW73}. We have assumed here, for simplicity, random motion of molecules along the $x$ axis only, in the case of a 3-d random motion of molecules, the only change in the final result would be that pressure will instead be given by $p=mn\overline{V_{\rm x}^2}=mn \overline{V^2}/3$ \cite{85}. 

The stress component $T'^{11}=-E^2/8\pi$ (Eq.~(\ref{eq:84.3a})) in the electromagnetic stress-energy tensor that gives rise to additional, electromagnetic momentum ${\mathbf P}_{\rm add}=-E^2\tau{\mathbf v}/(8\pi c^2)$ (Eq.~(\ref{eq:p31.2})), is, with an opposite sign, akin to the pressure in the ideal gas, that gives rise to the mechanical momentum, ${\mathbf P}_{\rm m}$ (Eq.~(\ref{eq:p31.2b1})). The energy-flux and momentum density exist within the moving medium   responsible for the force on the capacitor plates, in one case the ideal gas having molecular pressure and in the other case the electromagnetic fields possessing tension along the field lines. 

The mechanical momentum  ${\mathbf P}_{\rm m}=(p/ c^2){\mathbf v}\tau$ in the system, however, does not form a part of the {\em electromagnetic} momentum of the system, as the formulation for ${\mathbf P}_{\rm em}$ (Eqs.~(\ref{eq:84a.4}) or (\ref{eq:84a.7c})) involves only electromagnetic quantities (electric and magnetic fields).
The electromagnetic fields, resulting from the electromagnetic interactions between the constituents of the charged system, seem to possess all the information of the electromagnetic energy-momentum within the system, including any additional electromagnetic momentum contribution, ${\mathbf P}_{\rm add}$, from the electromagnetic forces within the moving system.  
The stabilization forces, keeping the system in equilibrium against these electromagnetic forces, make equal and opposite contributions to the additional momentum due to the electromagnetic forces. Since  electromagnetic fields could account for only the electromagnetic interactions but not any mechanical interactions that may exist in the system, the {\em mechanical momentum}, ${\mathbf P}_{\rm m}={\cal E}{\mathbf v}/c^{2}$, due to the stabilization forces would not show up in the {\em electromagnetic momentum} formulation.

Nonetheless, by adding the mechanical momentum, contributed by the stabilization forces, 
to the electromagnetic momentum (Eq.~(\ref{eq:p31.2a})), we get the total (electromagnetic + mechanical) momentum of the system as 
\begin{eqnarray}
	\label{eq:p31.2b}
	{\mathbf P}_{\rm t}&=& {\mathbf P}_{\rm A}+{\mathbf P}_{\rm add}+{\mathbf P}_{\rm m}
	=\frac{2\pi\sigma^{2}\tau}{c^{2}}{\mathbf v} - \frac{E^2\tau}{8\pi c^{2}} {\mathbf v}+\frac{p\tau}{c^{2}}{\mathbf v}\nonumber\\
&=&\frac{{\cal E}}{c^{2}}{\mathbf v}\,.
\end{eqnarray}
Thus, for a system of rest-frame energy ${\cal E}$, moving with velocity ${\mathbf v}$, it is the {\em total} momentum (electromagnetic + mechanical) that transforms as the component of a 4-vector, to yield ${\cal E}{\mathbf v}/{c^{2}}$ .
\section{Plate surfaces parallel to the direction of motion}
Here we take the rest frame $\cal K'$ of the charged capacitor to be moving in the lab frame $\cal K$  with a uniform non-relativistic velocity $\mathbf v$ ($v\ll c$), along the $x$-axis but with capacitor plates lying  in the x--z planes and the plate separation along the $y$-axis, 
\begin{figure}[t]
	\includegraphics[width=\columnwidth]{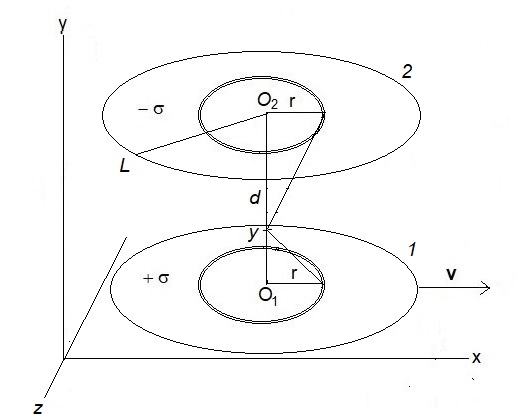}
	\caption{Computation of the scalar potential, ${\phi}$, at a point $y$ between the charged capacitor plates, due to circular rings of radii $r$, centered on $O_1$ and $O_2$, from the charged capacitor plates 1 and 2, lying in the x--z planes. The surface charge densities are $+\sigma$ and $-\sigma$, on plates 1 and 2 respectively. The capacitor system moves with velocity ${\bf v}$ with respect to the lab-frame.
\label{F3}}
\end{figure}

The scalar potential $\phi'$, in frame $\cal K'$, in this case a distance $y'$ from plate 1 (Fig.~{\ref{F3}), is  calculated as 
\begin{eqnarray}
\label{eq:84b.5}
 {\phi'} &=& \int_{0}^{L}\frac{\sigma 2\pi r\,{\rm d}r}{[y'^{2}+r^{2}]^{1/2}}
 -\int_{0}^{L} \frac{\sigma 2\pi r\,{\rm d}r}{[(d-y')^{2}+r^{2}]^{1/2}}\,,\nonumber\\
	 &&
\end{eqnarray}
which, for $y'\le d\ll L$, yields
\begin{eqnarray}
\label{eq:84b.6}
 {\phi'}  &=& \sigma 2\pi (d-2y') \,.
\end{eqnarray}
The vector potential ${\mathbf A}'=0$ in rest frame $\cal K'$, everywhere.

The electric field $\mathbf E'$ between the capacitor plates, in frame $\cal K'$, is then obtained as 
\begin{eqnarray}
\label{eq:84b.6a}
{\mathbf E'}&=& -\grad'{\phi'}  = 4\pi \sigma \hat{e}_{\rm y'}\,,
\end{eqnarray}
while the magnetic field is zero everywhere, in frame $\cal K'$.

We can compute the potentials in frame $\cal K$ from a Lorentz transformation of the 4-vector ($\phi,{\mathbf A}$) from  frame $\cal K'$ (Eq.~(\ref{eq:84b.6})) to $\cal K$, using (Eq.~(\ref{eq:84.1a1})). As $y'$ in frame $\cal K'$ corresponds to $y$ in  $\cal K$, we get  
\begin{eqnarray}
\label{eq:84b.6c}
{\phi}(y)  &=& \sigma 2\pi (d-2y)\nonumber\\
A_{\rm x}(y)&=&\sigma 2\pi (d-2y)\beta\,,
\end{eqnarray}
Here too the only finite component of vector potential ${\mathbf A}$ is along $x$-direction. Potentials $\phi$ and $A_{\rm x}$ depend only upon $y$ and are independent of $x$.

The electric field in frame $\cal K$, for $\gamma \approx 1$, is given by 
\begin{eqnarray}
\label{eq:84b.6d}
{\mathbf E}&=& {\mathbf E}'  = 4\pi \sigma \hat{\mathbf y}\,.
\end{eqnarray}
The electric field, ${\mathbf E}$, in frame $\cal K$ is thus along $y$-axis, and is constant between the capacitor plates. 

The magnetic field, ${\mathbf B}=\grad \times {\mathbf A}$, is then given by
\begin{eqnarray}
\label{eq:84b.6e}
{B}_{\rm z}&=&-\frac{\partial {A}_{\rm x}}{\partial y}  =  4\pi \sigma\beta\,.
\end{eqnarray}
which is consistent with ${\mathbf B}=\mbox{\boldmath $\beta$}\times {\mathbf E}$ \cite{PU85}. 

We can get the electromagnetic momentum, ${\mathbf p}_{\rm A}$,  as 
\begin{eqnarray}
	\label{eq:84b.9}
	{\mathbf P}_{\rm A}&=&\frac{ 2\pi \sigma^2\tau{\mathbf v}}{c^{2}}=\frac{E^2\tau {\mathbf v}}{8\pi c^{2}} \,,
\end{eqnarray}

The field momentum on the other hand is
\begin{eqnarray}
	\label{eq:84.21b}
	{\mathbf P}_{\rm f}&=&{\frac {1}{4\pi c}}\int(\mathbf {E} \times \mathbf {B}){{\rm d}\tau}=\frac{E^2\tau}{4\pi c^{2}}{\mathbf v}\,.
\end{eqnarray}

Here we see that the electromagnetic momentum (Eq.~(\ref{eq:84b.9})), calculated from the vector potential, is only half of that calculated from fields (Eq.~(\ref{eq:84.21b})). 

We can also obtain the electromagnetic momentum from the stress-energy tensor, which in the rest frame, $\cal K'$ is given by
\begin{eqnarray}
\label{eq:84.3d}
{\displaystyle T'^{\mu' \nu' }
 =\frac{E'^2}{8\pi}\operatorname {diag} (1,1,-1,1)}\:.
\end{eqnarray} 
Here $T'^{11}=E'^{2}/8\pi$ show a positive pressure along the $x$-axis, which is perpendicular to the field lines.

While  a Galilean transformation of the the energy-momentum tensor (\ref{eq:84.1c1}) leads to Eq.~(\ref{eq:84b.9}), a more appropriate Lorentz transformation for non-relativistic velocities (Eq.~(\ref{eq:84.3a1})) gives two additional, off-diagonal components for the stress-energy tensor in the lab frame $\cal K$ as
\begin{eqnarray}
\label{eq:84.3e}
T^{10} =T^{01} =(T'^{00} + T'^{11})\beta=\frac{2 E^2 \beta}{8\pi}\,,
\end{eqnarray} 
where in the lab frame $cT^{01}$ is the energy flux in $x$ direction
and $T^{10}/c$ is the $x$ component of the electromagnetic momentum density. 

That these stress-dependent terms are present in the electromagnetic momentum density expression ${\mathbf E} \times {\mathbf B}/4\pi c$ in frame ${\cal K}$ can be seen from 
\begin{eqnarray}
\label{eq:84a.33}
\frac{1}{4\pi c}{\mathbf E} \times {\mathbf B}&=&\frac{1}{4\pi c}{\mathbf E} \times (\mbox{\boldmath $\beta$}\times {\mathbf E})\nonumber\\ 
&=& \frac{1}{4\pi c}[\mbox{\boldmath $\beta$}E^2-{\mathbf E}(\mbox{\boldmath $\beta$} \cdot {\mathbf E})]=\frac{2 E^2\mbox{\boldmath $\beta$}}{8\pi c}\nonumber\\ 
&=&\frac{(T'^{00} + T'^{11})\beta}{c} \hat{\mathbf x}\:, 
\end{eqnarray}
where we have used $\mbox{\boldmath $\beta$} \cdot {\mathbf E}=0$ as ${\mathbf E}\perp\mbox{\boldmath $\beta$}$.

In Eqs.~(\ref{eq:84.3e}) or (\ref{eq:84a.33}), the term proportional to $T'^{00}\beta$, corresponds to ${\mathbf P}_{\rm A}$ (Eq.~(\ref{eq:84b.9})), the momentum calculated from the vector potential. The other term, proportional to $T'^{11}\beta$, corresponds to an additional energy flux as well as an additional momentum density. In consistency with Eq.~(\ref{eq:84.21b}), the net electromagnetic momentum, obtained from volume integral of $T^{10}/c$ in Eq.~(\ref{eq:84.3e}) is ${E^2\tau {\mathbf v}}/{4\pi c^{2}}$ since $T'^{11} = T'^{00}=E^2/{8\pi}$ here (Eq.~(\ref{eq:84.3d})).

Now here due to motion of the system, which is along the  $x$-axis, no work is done by the forces of mutual attraction between the plates, or by the mechanical forces keeping the plates separated, as they are along the $y$-axis. 
However, there are electromagnetic forces of repulsion along the plate surfaces, on similar charges {\em within} each plate, forces that might not be so apparent \cite{15}. 
There is a continuous energy flow taking place along the direction of motion, from the work done, by the forces of repulsion within each capacitor plate, which contributes to the momentum of the system moving along the $x$-axis.
Its evaluation, computationally rather involved \cite{15}, yields momentum contribution, ${\mathbf P}_{\rm add}=2\pi \sigma^2\tau{\mathbf v}/{c^{2}}$, which turns out to be the same as calculated from the stress component $T'^{11}$ ($=E^2/8\pi$) in the electromagnetic stress-energy tensor (Eq.~(\ref{eq:84.3d})), implying a momentum density $E^2{\mathbf v}/(8\pi c^2)$ (Eq.~(\ref{eq:84.3e})). 
This ${\mathbf P}_{\rm add}$ when added to ${\mathbf P}_{\rm A}$ in Eq.~(\ref{eq:84b.9}), gives the total electromagnetic momentum of the system
\begin{eqnarray}
	\label{eq:84.21d}
	{\mathbf P}_{\rm em}&=&{\mathbf P}_{\rm A}+{\mathbf P}_{\rm add}=\frac{ 2\pi \sigma^2\tau{\mathbf v}}{c^{2}}+\frac{E^2\tau {\mathbf v}}{8\pi c^{2}}=\frac{E^2\tau{\mathbf v}}{4\pi c^{2}}\,,\nonumber\\ 
&&
\end{eqnarray}
which agrees with the electromagnetic momentum computed from fields (Eq.~(\ref{eq:84.21b})).

There are, of course, stabilizing forces that keep the charges fixed in spite of the forces of repulsion between the charges on each plate and these contribute a mechanical momentum, ${\mathbf P}_{\rm m}=-{\mathbf P}_{\rm add}=-E^2\tau {\mathbf v}/(8\pi c^2)$, to the system. 
The {\em total} momentum of the system then is 
\begin{eqnarray}
	\label{eq:p31.2c}
	{\mathbf P}_{\rm t} &=& {\mathbf P}_{\rm A}+{\mathbf P}_{\rm add}+{\mathbf P}_{\rm m}=\frac{2\pi\sigma^{2}\tau}{c^{2}}{\mathbf v} + \frac{E^2\tau}{8\pi c^{2}} {\mathbf v}-\frac{E^2\tau}{8\pi c^{2}} {\mathbf v}\nonumber\\
&=&\frac{{\cal E}}{c^{2}}{\mathbf v}\,.
\end{eqnarray}
Thus, for a system of rest-frame energy ${\cal E}$, moving with velocity ${\mathbf v}$, it is the {\em total} momentum (electromagnetic + mechanical) that transforms as the component of a 4-vector, to yield ${\cal E}{\mathbf v}/{c^{2}}$. 

An argument has been put forward in the literature \cite{On22} that a proper explanation of the famous Trouton-Noble experiment \cite{59} lies in the fact that the electromagnetic momentum, ${\mathbf P}_{\rm A}$, of the system, calculated using the vector potential (Eqs.~(\ref{eq:84a.7}) and (\ref{eq:84b.9}), is independent of the plates orientations with respect to the direction of motion. However, this argument is not correct as we have shown here  that ${\mathbf P}_{\rm A}$ accounts for the electromagnetic momentum only partially, and that the {\em total }  electromagnetic momentum differs for different plate orientations. The real explanation of the Trouton-Noble experiment, and which has been shown explicitly \cite{31,58}, is that once the system is in equilibrium in the rest-frame  under the electromagnetic forces and the stabilization forces, it also remains in equilibrium in all frames of reference, since both types of forces will transform relativistically in exactly the same manner from one frame to another.

\section{Electromagnetic momentum of a charged sphere moving non-relativistically}
We consider a spherical shell of radius $\epsilon$ and a surface area $S=4\pi\epsilon^2$, with a uniform surface charge density $\sigma$, and thus of a total charge $e=4\pi\epsilon^2\sigma$.  The electric field of the charge in rest frame ${\cal K}'$ is ${\mathbf E}'= (e/{r^{2}})\hat{\mathbf r}$ for $r>\epsilon$, but zero  for $r<\epsilon$. From ${\mathbf r}=x \hat{\mathbf x}+y \hat{\mathbf y}+z \hat{\mathbf z}$, we can express ${\mathbf E}'= e{\mathbf r}/r^3=e (x \hat{\mathbf x}+y \hat{\mathbf y}+z \hat{\mathbf z})/{r^{3}}$ for $r>\epsilon$. The magnetic field of course is zero in ${\cal K}'$. Writing $\mbox{\boldmath $\beta$}={\mathbf v}/ c$ with  $\mbox{\boldmath 
$\beta$}= \beta \hat{\mathbf x}$, electromagnetic fields in lab frame ${\cal K}$ for a non-relativistic motion of the sphere, to a first order in $\beta$, with $\gamma \approx 1$, are given by ${\mathbf E}= {\mathbf E}'$ and ${\mathbf B}=\mbox{\boldmath $\beta$}\times {\mathbf E}$ \cite{PU85}.

\begin{figure}[t]
	\includegraphics[width=\columnwidth]{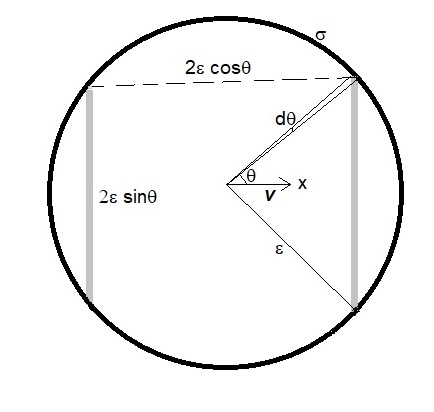}
	\caption{A spherical shell of radius $\epsilon$ with a uniform surface charge density $\sigma$, moving with velocity $\bf v$ along the x-axis. Shown in gray are two symmetrically placed circular rings, each of radius $\epsilon \sin \theta$ and of angular width $d\theta$, lying on two opposite sides of the spherical shell separated by a distance $l= 2\epsilon\cos \theta$ along the $x$-axis. 
 \label{F4}}
\end{figure}

The electromagnetic energy, ${\cal E}'$, of the system is the self-potential energy, evaluated from a double surface integral over the charged sphere, equal to a volume integral of the field energy density $E'^2/8\pi$ \cite{PU85}, 
\begin{eqnarray}
\label{eq:84b.4d}
{\cal E}' &=& \frac{1}{2} \int\int \frac{\sigma ({\mathbf x})\sigma ({\mathbf x}')}{|{\mathbf x}-{\mathbf x}'|}\,{\rm d}S\,{\rm d}S'
= \frac {1}{8\pi}\int_{\epsilon}^{\infty} E'^2 4\pi r^2{\rm d}r\nonumber\\
&=&\frac{e^2}{2\epsilon}=6\pi \sigma^2 \tau
\end{eqnarray}
where $\tau=4\pi \epsilon^{3}/3$ is the total volume of the sphere.

With the rest frame field energy ${\cal E}'$, one can associate an electromagnetic mass, ${\cal E}'/c^{2}$, and thereby, for a motion with velocity ${\mathbf v}$ in lab frame $\cal K$, an electromagnetic momentum 
\begin{eqnarray}
	\label{eq:84.21e}
	{\mathbf P}_{\rm A}={6\pi \sigma^2 \tau}\frac{\mathbf v}{c^{2}}=\frac{e^2}{2\epsilon}\frac{\mathbf v}{c^{2}}\,, 
\end{eqnarray}
using the vector potential (Eq.~(\ref{eq:84a.4c3})). This essentially results from the Lorentz transformation of the energy ${\cal E}'$ in rest frame $\cal K'$ which gives, for a non-relativistic motion ($\gamma \approx 1$), a momentum ${\cal E}' {\mathbf v}/{c^2}$ in lab frame $\cal K$.

However, from the electric and magnetic fields, $\mathbf E$ and $\mathbf B$, of a uniformly moving charge, from volume integral of momentum density (Eq.~(\ref{eq:84a.4})), one gets for the field momentum \cite{29} 
\begin{eqnarray}
\label{eq:84.31}
{\mathbf P}_{\rm f}=\int \frac{{\mathbf E} \times {\mathbf B}}{4\pi c}\,{\rm d}\tau=\frac{4}{3}\left(\frac{e^{2}}{2\epsilon}\right) \frac{\mathbf v}{c^{2}}\:,
\end{eqnarray}
which is higher than ${\mathbf P}_{\rm A}$ by a factor of $4/3$, which has been a source of confusion for long \cite{1,2,29}.

In order to comprehend the occurrence of this extra factor of 4/3 in the field momentum, we 
investigate various terms in momentum density  emerging out of the stress energy tensor for the moving charged sphere in the lab frame ${\cal K}$. 
For this, we first write $T'^{\mu' \nu'}$ in the rest frame ${\cal K}'$, where energy flux and momentum density are zero but the Maxwell stress tensor may have non-zero terms. Employing Eqs.~(\ref{eq:84a.24}), (\ref{eq:84a.25}) and (\ref{eq:84a.26}), we can write  (for $r>\epsilon$)
\begin{eqnarray}
\label{eq:84.3g}
{\displaystyle T'^{\mu' \nu'}
{\displaystyle =\frac{e^2}{8\pi r^6}{\begin{bmatrix}r^2&0&0&0\\0&r^2-2x^2&-2xy&-2xz\\0&-2yx&r^2-2y^2&-2yz\\0&-2zx&-2zy&r^2-2z^2\end{bmatrix}}}}. 
\end{eqnarray} 

Though a Galilean transformation of the the energy-momentum tensor (\ref{eq:84.1c1}) does yield  Eq.~(\ref{eq:84.21e}), a more appropriate Lorentz transformation for non-relativistic velocities (Eq.~(\ref{eq:84.3a1})) gives rise to the following additional non-vanishing components of the stress-energy tensor in the lab frame $\cal K$ 
\begin{eqnarray}
\label{eq:84.3i}
T^{10} &=& T^{01}=(T'^{00} + T'^{11})\beta=\frac{e^2(r^2-x^2)}{4\pi r^6}\beta\nonumber\,,\\
T^{20} &=& T^{02}= T'^{12}\beta=\frac{-e^2 xy}{4\pi r^6}\beta\nonumber\,,\\
T^{30} &=& T^{03}=T'^{13}\beta=\frac{-e^2 xz}{4\pi r^6}\beta\,,
\end{eqnarray} 
where terms of order $\beta^2$ or higher have been dropped. All the non-vanishing components of the stress-energy tensor in the rest frame ${\cal K}'$ (Eq.~(\ref{eq:84.3g})) remain unchanged to that order, in the lab frame ${\cal K}$.
However, there are additional terms in energy flux $cT^{0i}$ and momentum density $T^{i0}/c$, proportional to stress terms, $T'^{1i'}$, in fields, apart from the term proportional to the energy density $T'^{00}$.  
Curiously there are momentum density terms $T^{20}$ and $T^{30}$, which are finite along $y$ and $z$ axes even though the system has motion only along the $x$ axis.

We can look for the presence of stress-dependent terms also in the electromagnetic momentum density expression ${\mathbf E} \times {\mathbf B}/4\pi c$, by writing  
\begin{eqnarray}
\label{eq:84.32}
\frac{1}{4\pi c}{\mathbf E} \times {\mathbf B}\!\!&=&\!\!\frac{1}{4\pi c}{\mathbf E} \times (\mbox{\boldmath $\beta$}\times {\mathbf E})\nonumber\\ 
&=&\!\! \frac{1}{4\pi c}[\mbox{\boldmath $\beta$}E^2-{\mathbf E}(\mbox{\boldmath $\beta$} \cdot {\mathbf E})]\nonumber\\ 
&=&\!\!\frac{e^{2}\beta}{4\pi c r^6}[(r^2-x^2) \hat{\mathbf x}- xy\hat{\mathbf y}- xz \hat{\mathbf z}]. 
\end{eqnarray}

A comparison of Eqs.~(\ref{eq:84.32}) with (\ref{eq:84.3i}) shows the role played by the Maxwell stress tensor terms in the electromagnetic momentum of a moving system. There is one to one  correspondence of terms in the two expressions, all terms arising from stress in Eq.~(\ref{eq:84.3i}) are present in Eq.~(\ref{eq:84.32}) and vice versa.

From an average over a spherical surface of radius $r$, we have $\overline{{x}^2}={r^2}/3$, $\overline{{xy}}=0$, $\overline{{xz}}=0$. Then a volume integral of the momentum density $T^{i0}/c$ from Eq.~(\ref{eq:84.3i}) yields the only non-zero value for 
\begin{eqnarray}
\label{eq:84.33}
\frac{1}{c}\int T^{10}\,{\rm d}\tau &=&\frac{1}{c}\int(T'^{00} + T'^{11})\beta\,{\rm d}\tau\nonumber\\
&=& 
\int\frac{e^2r^2}{8\pi c r^6}\beta\,{\rm d}\tau+\int\frac{e^2(r^2-2x^2)}{8\pi c r^6}\beta\,{\rm d}\tau\nonumber\\
&=&\frac{e^{2}\beta}{2\epsilon c}+\frac{1}{3}\frac{e^{2}\beta}{2\epsilon c},
\end{eqnarray}
which is along $x$-axis, the direction of motion.

Then we have the electromagnetic momentum ${\mathbf P}_{\rm em}={\mathbf P}_{\rm f}={{4\cal E}\mathbf v}/{3c^{2}}$,
where, the electromagnetic momentum, apart from the term ${\cal E}{\mathbf v}/{c^{2}}$  coming from the energy density $T'^{00}$   
and equal to ${\mathbf P}_{\rm A}$, determined using the vector potential (Eq.~(\ref{eq:84.21e})), 
possesses an additional term, ${\cal E}{\mathbf v}/3{c^{2}}$, due to stress in the fields.
Thus ${\mathbf P}_{\rm f}$ in Eq.~(\ref{eq:84.31}) comprises an additional contribution ${\mathbf P}_{\rm add}={\cal E}{\mathbf v}/3c^{2}$ from stress in the fields. 

From a physical perspective, this additional momentum term, ${\mathbf P}_{\rm add}$, arises from an energy flow associated with the work done by the self-fields of the moving charged sphere and has been calculated for a relativistically moving system \cite{15,20,Mo11}. We show here that even for a non-relativistic motion of the system ${\mathbf P}_{\rm add}$ could be significant. 

Due to its velocity $\mathbf v$, the charged spherical shell of surface thickness, say $\delta$, and thus of a volume charge density $\sigma/\delta$ inside the thin shell, carries a current density ${\mathbf j}= \sigma {\mathbf v}/\delta$. Then the electric self-field ${\mathbf E}= 2\pi \sigma \hat{\mathbf r}$ within the shell, which is an average of the field just outside and inside the sphere \cite{PU85}, does work at a temporal rate, $ {\mathbf j} \cdot {\mathbf E}$  per unit volume, or $ {\mathbf j} \cdot {\mathbf E}\:\delta=2\pi \sigma^2 v\cos \theta $ on a unit surface area of the shell. 

We consider, now, two infinitesimal surface elements in the shape of circular rings, each of radius $\epsilon \sin \theta$ and angular width $d\theta $, and thus of surface area ${\rm d}S=2\pi \epsilon^{2}\sin \theta\:{\rm d}\theta$ on two opposite sides of the spherical shell, separated by a distance $l= 2\epsilon\cos \theta$ along the direction of motion (Fig.~{\ref{F4}). 
The rate of work being done by the self-field of the sphere on the left ring, accordingly, is $-2\pi \sigma ^{2} v\cos \theta\:{\rm d}S$ while on the right ring it is $2\pi \sigma ^{2}v\,\cos \theta\:{\rm d}S$. The right ring thus gains energy at the cost of the left ring, implying 
a transport of energy $2\pi \sigma ^{2} \,{\mathbf v}\,\cos \theta\:{\rm d}S$ across a distance $l=2\epsilon\,\cos\theta$, per unit time.  

This flow of energy, in turn, implies a momentum 
$2\pi \sigma ^{2}  {\mathbf v}\cos \theta\:{\rm d}S\,l/c^2=4\pi \sigma ^{2} \,\epsilon {\mathbf v}\cos^{2} \theta\,{\rm d}S/c^2$, 
within a cross-section ${\rm d}S$ of the moving sphere.

An integration of momentum, due to its electric self-fields, over the total cross-section of the moving charged sphere, yields an additional contribution 
\begin{eqnarray}
\label{eq:84b.4e}
{\mathbf P}_{\rm add}&=&\frac{4\pi \sigma^{2}{\mathbf v}} {c^{2}}\int^{\pi/2}_{0} \epsilon\cos^{2}\theta\: 2\pi \epsilon^{2}\sin \theta\:{\rm d}\theta\nonumber\\
&=&\frac{2\pi \sigma^{2}\tau {\mathbf v}} {c^{2}}.
\end{eqnarray}

The total electromagnetic momentum of the moving charged spherical system therefore is 
\begin{eqnarray}
\label{eq:84b.4f}
{\mathbf P}_{\rm em}&=&{\mathbf P}_{\rm A}+{\mathbf P}_{\rm add}
= 6\pi \sigma ^{2} \tau\,\frac{\mathbf v}{c^{2}}+2\pi \sigma ^{2} \tau\,\frac{\mathbf v}{c^{2}}
\nonumber\\
&=&\frac{4}{3}\left(\frac{e^{2}}{2\epsilon}\right) \frac{\mathbf v}{c^{2}}\,,
\end{eqnarray}
which 
thus explains the famous factor of 4/3 in the electromagnetic momentum of a moving charged particle, when work done by self-fields is taken into account, even for non-relativistic velocities \cite{1,2,29}.

If we consider the mechanical momentum  ${\mathbf P}_{\rm m}=-e^{2} {\mathbf v}/(6\epsilon c^2)$, due to stabilizing forces, equal and opposite to forces of self-repulsion, that keep the charges `glued' on the spherical surface, in spite of the forces of repulsion between the charges, then total momentum of the system becomes 
\begin{equation}
	\label{eq:p31.2d}
	{\mathbf P}_{\rm t}= {\mathbf P}_{\rm em}+{\mathbf P}_{\rm m}=\frac{4}{3}\frac{e^{2}}{2\epsilon} {\mathbf v}-\frac{1}{3}\frac{e^{2}}{2\epsilon} \frac{\mathbf v}{c^{2}}=\frac{e^{2}}{2\epsilon} \frac{\mathbf v}{c^{2}}\,.
\end{equation}
Thus, for a charged sphere of rest-frame electromagnetic energy ${\cal E}=e^{2}/2\epsilon$, moving with velocity ${\mathbf v}$, it is the {\em total} momentum (electromagnetic + mechanical) that transforms as the component of a 4-vector, to yield $(e^{2}/2\epsilon)\:{\mathbf v}/ c^{2}$ .

A question that could justifiably be raised here is where is the need to go through the onerous  task of determining ${\mathbf P}_{\rm add}$ and ${\mathbf P}_{\rm m}$, when {\em we know} that the two, being equal and opposite, are finally going to get cancelled from the total momentum of the system. Would it not be rather much simpler to drop them altogether and stick to the momentum derived from a Lorentz transformation of the rest frame energy ${\cal E}'$ alone. 
Such a move, however, would mean a modification in the formulation for the electromagnetic momentum density (Eq.~(\ref{eq:84.32})) which, as we have seen, encompasses stress terms (Eq.~(\ref{eq:84.3i})) from the Maxwell stress tensor, giving rise to the additional electromagnetic momentum terms for a moving charged system. Actually such a modification would imply that under a Lorentz transformation the energy and momentum densities of a system, in place of being components of the stress-energy tensor, energy-momentum behaving instead as a 4-vector.

In fact, the appearance of extra factor of 4/3 in the electromagnetic momentum of a moving charged sphere had caused so much confusion over such a long period (about a century) that it even led to serious proposals for such modifications \cite{Rohrlich60} in the standard definition of electromagnetic field energy-momentum, modifications that appeared even in standard text-books \cite{1,2}. However, it has subsequently been shown that there is no real need for such modifications in the standard formulation of electromagnetic field energy-momentum \cite{31}. 
Instead, as we have shown here, for 
a moving charged system the problem of the apparently anomalous electromagnetic momentum is successfully resolved when one realizes that in the lab frame, during a Lorentz transformation, there could be additional contribution  arising from the finite Maxwell stress tensor terms in the rest frame. 

In hindsight, it can be said that all one requires is that in an  electromagnetic system a proper distinction be maintained between the {\em electromagnetic} momentum and the {\em total} momentum, the latter including all non-electromagnetic momentum contributions as well. Thus, while ${\mathbf P}_{\rm add}$ forms part of the electromagnetic momentum, the non-electromagnetic momentum ${\mathbf P}_{\rm m}$ does not. In the total momentum of the system even the stabilizing forces (Poincar\'{e} stresses \cite{34}), which may be non-electromagnetic, are taken into account which in fact, cancel the Maxwell stress tensor terms, thereby making the total energy-momentum of the system a 4-vector 
Thus, taking each of these momentum contributions in proper account is the key to understand why the standard formulation of electromagnetic field energy-momentum does not require any modifications.

Further, there is no need for any modifications in the standard formulation of electromagnetic field energy-momentum to eliminate the factor of 4/3, as this `enigmatic' factor, in reality, is not something to do with electromagnetic properties of a spherical {\em charged} system. It arises because of the stress/pressure in the system, irrespective of whether its genesis is of electromagnetic origin or caused by something else. For instance, even in a non-electromagnetic  system like a moving perfect fluid system, pressure makes a similar contribution to the momentum of the system \cite{84}. In fact, in the case of radiation, where pressure $p$ is related  to energy density $u$ of the fluid as $p= u/3$, the momentum density of the system is $(4/3)\: u\: v/c^2$  and the self-repulsion force on a charged spherical surface can be considered mathematically equivalent to a sphere filled with electromagnetic radiation, with a uniform outward pressure, and one then arrives at a factor of 4/3 in the momentum of this radiation filled sphere \cite{84}, having no electric charges and no consequential electromagnetic fields therefrom. What matters there is the presence of pressure in the system that yields, in the presence of a bulk motion of the system, terms in energy-momentum of the system which are of the similar nature as in electromagnetic systems like a moving spherical charge or a charged spherical capacitor system \cite{Si24}.

Historically, it was in the {\em classical electron model} where the enigmatic factor of 4/3 had appeared in the electromagnetic momentum calculations, however, the above formulation is applicable to any finite spherical charge distribution that is undergoing a uniform motion. The results for the charged sphere were derived here for a non-relativistic motion, so as to bring out the relation between various quantities like work done, energy flow and the consequential momentum in the system, in a simplest possible manner. A brief account of the relevant history along with a full relativistic treatment, where contribution by the self-forces of the system between various constituents of the charged system during Lorentz contraction also plays a crucial role, for instance, the work done during the Lorentz contraction which turns a sphere into a spheroid due to motion, can be found in the literature \cite{15}. The derivation of energy-momentum of a relativistically moving system from the stress-energy tensor, is presented here in Appendix A, results of course consistent with those derived from the self-forces in the system. 
\section{Conclusions}
From an analysis of the momentum in various electrically charged systems it was shown that the momentum, determined using the vector potential, does not account for the total electromagnetic momentum of a moving system. The momentum thus obtained was shown actually to be the same as derived from a Lorentz transformation of the rest-frame electromagnetic energy of the system, assuming the electromagnetic energy-momentum to be a 4-vector. The energy-momentum densities of electromagnetic fields, however, form components of the stress-energy tensor, and their transformation from rest frame to lab frame includes contributions from the Maxwell stress tensor terms. It was demonstrated that the genesis of these additional terms from stress in the electromagnetic fields, could be traced, from a physical perspective, to electromagnetic forces between various constituents of a moving system contributing to electromagnetic momentum, contributions that might not always be very obvious. These subtle, additional contributions to the electromagnetic momentum could be significant even for a non-relativistic motion. It was shown that the work being done by the electromagnetic forces on opponent parts of the system results in a transport of electromagnetic energy in the system, contributing to the electromagnetic momentum. The total momentum of the system encompasses also a contribution of mechanical momentum. due to the non-electromagnetic forces of stabilization within the moving system, which however, could not have been represented in the electromagnetic momentum since the latter can account for only the electromagnetic interactions in the system. 
\section*{Data Availability}
Data sharing not applicable to this article as no datasets were generated or analysed during the current study.
\section{Disclosures and declarations}
The author has no conflicts of interest/competing interests to declare that are relevant to the content of this article. No funds, grants, or other support of any kind was received from anywhere for this research.
\appendix

\section{Energy-momentum of charged systems moving relativistically}
We assume the rest frame $\cal K'$ of the charged system, with respect to lab frame $\cal K$, moving relativistically along the $x$-axis with a uniform velocity $\beta$ and the Lorentz factor $\gamma =1/\sqrt{1-\beta^{2}}$. We want to determine energy-momentum  of the system at some instant, say $t=0$, in $\cal K$, implying $x'=\gamma x, y'=y, z'=z$.  
\subsection{Capacitor moving normal to plate surfaces}
The electric field in the lab frame $\cal K$, is ${E}_{\rm x}= {E}'$, while the magnetic field,is zero. The volume is ${\rm d}\tau={\rm d}\tau'/\gamma$.

From Eqs.~(\ref{eq:84.3a11}) and (\ref{eq:84.3a}), we get in the lab frame 
\begin{eqnarray}
\label{eq:84a.3e1}
T^{\mu \nu} &=&
{\displaystyle \frac{E'^2}{8\pi}{\begin{bmatrix}\gamma^2(1-\beta^2) &\gamma^2(1-1)\beta & 0 & 0\\\gamma^2(1-1)\beta & \gamma^2(\beta^2-1) & 0 &0\\0&0&1&\\0&0&0&1\end{bmatrix}}}
\nonumber\\
&=&
\frac{E'^2}{8\pi}\operatorname {diag} (1,-1,1,1)=T'^{\mu' \nu'}\:.
\end{eqnarray} 

Electromagnetic  energy and momentum of the system in the lab frame are obtained from the volume integrals
\begin{eqnarray}
	\label{eq:84.22c}
	{\cal E}&=&\int {T^{00}} {{\rm d}\tau}={\frac {1}{8\pi}}\int {E'^2} \frac{{\rm d}\tau'}{\gamma}=\frac{{\cal E'}}{\gamma}\,,\\
	\label{eq:84.22b}
	{\mathbf P}_{\rm em}&=&{\frac {1}{c}}\int {T^{10}} {{\rm d}\tau}=0\,,
\end{eqnarray}
consistent with those derived from the self-forces \cite{15}.

\subsection{Capacitor moving parallel to plate surfaces}
The electric field for a relativistic motion, in the lab frame $\cal K$, is ${E}_{\rm y}= \gamma{E}'$, while the magnetic field,is ${\mathbf B}=\mbox{\boldmath $\beta$}\times {\mathbf E}$ or ${B}_{\rm z}= \gamma\beta{E}'$.
From Eqs.~(\ref{eq:84.3a11}) and (\ref{eq:84.3d}), we get 
\begin{eqnarray}
\label{eq:84.3e2}
T^{\mu \nu} &=& {\displaystyle \frac{E'^2}{8\pi}{\begin{bmatrix}\gamma^2(1+\beta^2) &2\gamma^2\beta & 0 & 0\\2\gamma^2\beta & \gamma^2(\beta^2+1) & 0 & 0\\
0&0&-1&\\0&0&0&1\end{bmatrix}}}\,,\nonumber\\
	&&
\end{eqnarray} 

Energy and momentum of the system are 
\begin{eqnarray}
	\label{eq:84.23c}
	{\cal E}&=&\int {T^{00}} {{\rm d}\tau}=\int\frac{E'^2}{8\pi} \gamma^2(1+\beta^2) \frac{{\rm d}\tau'}{\gamma}\nonumber\\
	&=&\gamma{\cal E'}(1+\beta^2)\,,\\
	\label{eq:84.23b}
	{\mathbf P}_{\rm em}&=&{\frac {1}{c}}\int {T^{10}} {{\rm d}\tau}
	=\frac {2\gamma{\cal E'}\mbox{\boldmath $\beta$}}{c}\,,
\end{eqnarray}
in consistency with the results obtained from self-forces \cite{15}.
\subsection{A charged sphere moving relativistically}
The electromagnetic fields in lab frame $\cal K$ are given by ${E}_{\rm x}={E}'_{\rm x}$, ${E}_{\rm y}=\gamma{E}'_{\rm y}$, ${E}_{\rm z}=\gamma{E}'_{\rm z}$ and ${\mathbf B}=\mbox{\boldmath $\beta$}\times {\mathbf E}$, where ${\mathbf E}'= e (x' \hat{\mathbf x}+y' \hat{\mathbf y}+z' \hat{\mathbf z})/{r'^{3}}$ is the field in rest frame $\cal K'$.

From Eqs.~(\ref{eq:84.3a11}) and (\ref{eq:84.3g}), we get the energy and momentum densities in the lab frame as 
\begin{eqnarray}
\label{eq:84.3k}
T^{00} &=&\frac{e^2\gamma^2[r'^2+(r'^2-2x'^2)\beta^2]}{8\pi r'^6}\,,\\
\label{eq:84.3l}
T^{10}/c &=&\frac{e^2(r'^2-x'^2)}{4\pi r'^6}\gamma^2\frac{\beta}{c}\nonumber\,,\\
T^{20}/c &=&\frac{-e^2 x'y'}{4\pi r'^6}\gamma\frac{\beta}{c}\nonumber\,,\\
T^{30}/c &=&\frac{-e^2 x'z'}{4\pi r'^6}\gamma\frac{\beta}{c}\,.
\end{eqnarray} 
A volume integral of, say, energy density in $\cal K$ is
\begin{eqnarray}
	\label{eq:84.21f}
	{\cal E}&=&\int T^{00} {{\rm d}\tau}= \frac{e^2\gamma^2}{8\pi r'^6}\int[r'^2+(r'^2-2x'^2)\beta^2]\frac{{\rm d}\tau'}{\gamma}\,.\nonumber\\
	&&
\end{eqnarray}
Thus we get for the energy and momentum of the system
\begin{eqnarray}
\label{eq:84.3n}
{\cal E}&=&\frac{e^{2}\gamma}{2\epsilon} \left(1+\frac{\beta^2}{3}\right)=\gamma{\cal E}'\left(1+\frac{\beta^2}{3}\right)\,,\\
\label{eq:84.3m}
{\mathbf P}_{\rm em}&=&{\frac {1}{c}}\int {T^{10}} {{\rm d}\tau}=\frac{4}{3}\frac{e^{2}}{2\epsilon}\frac{\gamma\mbox{\boldmath $\beta$}}{c}=\frac{4}{3}{\cal E}' \frac{\gamma\mbox{\boldmath $\beta$}}{c}\,.
\end{eqnarray} 
Electromagnetic energy and momentum computed thus 
are the same as derived from the self-forces \cite{15}.
{}

\begin{thebibliography}{00}
\bibitem{15} A. K. Singal, J. Phys. A {vol. 25}, pp. 1605-1620 (1992).
\bibitem{31}  A. K. Singal, Am. J. Phys. {vol. 84}, pp. 780-785 (2016).
\bibitem{On22} V. Onoochin, Found. Phys. {vol. 52}, 35 (2022).
\bibitem{84}  A. K. Singal, Found. Phys. {vol. 51}, 4 (2021).
\bibitem{85} A. K. Singal,  Eur. J. Phys. {vol. 41}, 065602 (2020).
\bibitem{MTW73} C. W. Misner, K. S. Thorne and J. A, Wheeler, {\em Gravitation} (Freeman, San Fransisco, 1973).
\bibitem{SC85} B.F. Schutz, {\em A first course in general relativity} (Cambridge University, Cambridge, 1985). 
\bibitem{RI06} W. Rindler, {\em Relativity - Special, General and Cosmological} 2nd ed. (Oxford University Press, Oxford, 2006).
\bibitem{1} J. D. Jackson, {\em Classical electrodynamics} 2nd ed. (Wiley, New York, 1975).
\bibitem{2} W. K. H. Panofsky and M. Phillips, {\it Classical electricity and magnetism} 2nd ed. (Addison-Wesley, Massachusetts, 1962).
\bibitem{29} R. P. Feynman, R. B. Leighton and M. Sands,{\em The Feynman Lectures on Physics}, Vol. II, Addison-Wesley, Mass (1964).
\bibitem{Go50} H. Goldstein, {\em  Classical Mechanics}, (Addison-Wesley, Reading, 1950).
\bibitem{Da20} A. Davis and V. Onoochin, PIER L {vol. 94}, 151 (2020).
\bibitem{Ma65} J. C. Maxwell, ``A Dynamical Theory of the Electromagnetic Field.'' Phil. Trans. Roy. Soc. London {vol. 155}, pp. 459-512 (1865).
\bibitem{Ko78}  E. J. Konopinski, Am. J. Phys. {vol. 46}, pp. 499-502 (1978).
\bibitem{Se98}  M. D. Semon and J. R. Taylor, Am. J. Phys. {vol. 64}, pp. 1361-1369 (1996).
\bibitem{Gr12} D. J. Griffiths, Am. J. Phys. {vol. 80}, pp. 7-18 (2012).
\bibitem{Es18} H. Ess\'{e}n, Eur. J. Phys. {vol. 39}, 025202 (2018).
\bibitem{Lo04} H. A. Lorentz {\em Proc. Acad. Sci. Amst.} {vol. 6}, 809 (1904). 
Reprinted in: H. A. Lorentz, A. Einstein, H. Minkowski and H. Weyl, {\em The principle of relativity}, p. 9 (Dover, New York, 1952).
\bibitem{Ei05} A. Einstein, {\em Ann. Physik} {vol. 17}, 891 (1905). Reprinted in:  H. A. Lorentz, A. Einstein, H. Minkowski and H. Weyl, {\em The principle of relativity}, p. 37 (Dover, New York, 1952).
\bibitem{PU85} E. M. Purcell, {\em Electricity and Magnetism  - Berkeley Phys. Course vol. 2} 2nd ed. (McGraw-Hill, New York, 1985).
\bibitem{34} H. Poincar\'{e}, Rend. Circ. Mat. Palermo {21} 129 (1906). English trans. with modern notation in H. M. Schwartz, Am. J. Phys. {40} pp. 862-872 (1972)
\bibitem{59}  F. T. Trouton and H. R. Noble,  Phil. Trans. Roy. Soc. London A {vol. 202}, pp. 165-181 (1903)
\bibitem{58} A. K. Singal, Am. J. Phys. {vol. 61}, pp. 428-433 (1993).
\bibitem{20} A. D. Yaghjian, {\it Relativistic Dynamics of a charged sphere}, 2nd ed. (Springer, New York, 2006).
\bibitem{Mo11} V. B. Morozov, Phys.-Usp. {vol. 54}, 371 (2011).
\bibitem{Rohrlich60} F. Rohrlich, Am. J. Phys. {vol. 28}, pp. 639-643 (1960).
\bibitem{Si24} A. K. Singal, Eur. J. Phys. {\bf 45}, 065202 (2024).
\end{thebibliography}
\end{document}